\documentclass[12pt, peerreview , onecolumn]{IEEEtran}

\usepackage{tikz}
\usepackage[sumlimits,intlimits]{amsmath}
\usepackage{amsmath,amsfonts,amssymb}%
\usepackage{bm}%
\usepackage{graphicx,graphics}%
\usepackage{epstopdf}
\usepackage{cases}%
\usepackage[noadjust]{cite}%
\usepackage{color}%
\usepackage{cite,url}
\usepackage{verbatim}
\usepackage{algorithm}
\usepackage{algorithmic}
\usepackage{balance}
\usepackage{multirow}
\usepackage{multicol}
\usepackage{stfloats}
\usepackage{subfigure}
\usepackage{xspace}
\usepackage{enumerate}
\usepackage{rotating}
\usepackage[numbers,sort&compress]{natbib}

\ifCLASSOPTIONpeerreview

\fi

\begin{document}

\ifCLASSOPTIONpeerreview
\title{\hspace*{-12pt}{Channel, Phase Noise, and Frequency Offset in \\ \vspace*{-10pt}
    OFDM Systems: Joint Estimation, Data Detec-\\ \vspace*{-10pt}
    tion, and Hybrid Cram\'{e}r-Rao Lower Bound}}
\fi

\author{\IEEEauthorblockN{Omar H. Salim, \textit{Student Member, IEEE}, Ali A.~Nasir, \textit{Member, IEEE}, Hani Mehrpouyan, \textit{Member, IEEE}}, Wei Xiang, \textit{Senior Member, IEEE}, Salman Durrani, \textit{Senior Member, IEEE}, and Rodney A. Kennedy, \textit{Fellow, IEEE}}

\maketitle

{\let\thefootnote\relax\footnotetext{ Omar H. Salim and Wei Xiang are with the School of Mechanical and Electrical Engineering, University of Southern Queensland, Australia. Ali A. Nasir, Salman Durrani, and Rodney Kennedy are with Research School of Engineering, The Australian National University, Canberra, Australia. Hani Mehrpouyan is with the Department of Electrical and Computer Engineering, California Sate University, Bakersfield. Emails: OmarHazim.Salim@usq.edu.au, ali.nasir@anu.edu.au, hani.mehr@ieee.org, wei.xiang@usq.edu.au, salman.durrani@anu.edu.au, and rodney.kennedy@anu.edu.au. This research was partially supported under Australian Research Council's Discovery Projects funding scheme (project number DP140101133). \vspace{-00pt}}}

\vspace{-12pt}

\begin{abstract}
Oscillator phase noise (PHN) and carrier frequency offset (CFO) can adversely impact the performance of orthogonal frequency division multiplexing (OFDM) systems, since they can result in inter carrier interference and rotation of the signal constellation. In this paper, we propose an expectation conditional maximization (ECM) based algorithm for joint estimation of channel, PHN, and CFO in OFDM systems. We present the signal model for the estimation problem and derive the hybrid Cram\'{e}r-Rao lower bound (HCRB) for the joint estimation problem. Next, we propose an iterative receiver based on an \emph{extended Kalman filter} for joint data detection and PHN tracking. Numerical results show that, compared to existing algorithms, the performance of the proposed ECM-based estimator is closer to the derived HCRB and outperforms the existing estimation algorithms at moderate-to-high signal-to-noise ratio (SNR). In addition, the combined estimation algorithm and iterative receiver are more computationally efficient than existing algorithms and result in improved average uncoded and coded bit error rate (BER) performance.
\end{abstract}

\ifCLASSOPTIONpeerreview
    \newpage
\fi

\section{Introduction}
\subsection{Motivation and Literature Survey}
Orthogonal frequency division multiplexing (OFDM) is a powerful multi-carrier modulation technique for increasing the bandwidth efficiency of wireless communication systems. By converting a frequency-selective channel into multiple frequency-flat subchannels, OFDM can mitigate the detrimental effects of frequency-selective fading \cite{Nuaymi-2007wiley,Zhang-2008wiley}. Hence, OFDM has been adopted by existing and future wireless local area network (WLAN) standards such as IEEE $802.11$ac and IEEE $802.11$ad \cite{Paul-2008FQ,Zhu-2011May}. However, OFDM systems are much more sensitive to synchronization errors than single-carrier systems.

In practice, OFDM systems require timing offset estimation, carrier frequency offset (CFO) estimation, phase noise (PHN) tracking as well as channel estimation. Timing synchronization for OFDM systems has been well investigated for the past two decades \cite{{Chevillat-1987Sep},{Schmidl-1997Dec}}. Compared to timing offsets, OFDM is very sensitive to CFO and PHN, which arise due to instabilities and the thermal noise in the local oscillator \cite{article-PHASE-N-MODEL-I}, respectively. CFO and time varying PHN result in a common phase error (CPE) and inter-carrier interference (ICI) at the receiver, degrading the performance of OFDM systems  \cite{Lin-2006sept,Armada-1998Jan,Armada-2001Jun,Petrovi1-2003,Petrovi2-2003,Petrovic-2007Aug}. In particular, the impact of PHN in systems operating at higher carrier frequencies, e.g., V-band/$60$ GHz and E-band/$70$--$80$ GHz, can be even more profound \cite{article_mehr-comm-13}. Thus, as wireless communication systems and standards, e.g., IEEE $802.11$ad, migrate to millimeter-wave frequencies to take advantage of the large bandwidth in this band and adopt higher order modulations and closely spaced sub-carriers to achieve higher spectral efficiencies, it is increasingly important to develop efficient and accurate estimation and detection algorithms for compensating the effect of CFO and PHN in OFDM systems.

In order to jointly estimate channel, CFO, and time varying PHN, training signals are used in OFDM systems. In the context of point-to-point systems, joint channel and CFO estimation based on the expectation-maximization (EM) approach was proposed in \cite{Lee-2006Jan}. However, in \cite{Lee-2006Jan}, the authors do not take the effect of PHN into account. In \cite{Lin-2006sept} and \cite{Tao-2009Oct}, a MAP estimator was used for joint estimation of channel, CFO, and PHN. However, the estimation approach in \cite{Lin-2006sept} and \cite{Tao-2009Oct} is based on a small angle approximation (single-order Taylor series expansion of PHN), that adversely affects the performance of the estimation and data detection algorithms, especially for higher order modulations. In addition, as shown in this paper, the approach in \cite{Lin-2006sept} and \cite{Tao-2009Oct} can be computationally very complex. Recently, the authors in \cite{Carvajal-2013Jan} proposed a joint channel, CFO, and PHN estimation algorithm based on the sequential Monte Carlo and EM approaches. However, as explained in \cite{Septier-2008Aug}, the estimation complexity of Monte Carlo based method in \cite{Carvajal-2013Jan} is very high. More importantly, in \cite{Lin-2006sept,Tao-2009Oct,Carvajal-2013Jan}, the hybrid Cram\'{e}r-Rao lower bound (HCRB) for the joint estimation of channel impulse response (CIR), PHN, and CFO in OFDM systems is not derived and the performances of the proposed estimators are not benchmarked against their respective estimation performance bounds. Recently, the problem of joint estimation of channel, CFO, and PHN was considered in the context of OFDM relay networks in \cite{article-rui-sp-2013}. However, the approach in \cite{article-rui-sp-2013} is also based on the \emph{maximum a posteriori} (MAP) criterion, which is computationally very complex.

Given the time-varying nature of PHN, it needs to be tracked not only during the training interval but also during the data transmission interval. Hence, following the training period, a receiver structure for joint data detection and PHN mitigation in the data transmission period is required. In the existing literature, joint data detection and PHN mitigation was analyzed in \cite{Lin-2007May,Gong-2008May,Petrovic-2007Aug,Rabiei-20010Nov}. However, the algorithms proposed in \cite{Petrovic-2007Aug} and \cite{Gong-2008May} are based on the assumption of perfect knowledge of channel and CFO. Moreover, the PHN tracking and data detection approach presented in \cite{Lin-2007May} is computationally complex and suffers from performance degradation for higher order modulations. The PHN tracking in \cite{Rabiei-20010Nov} requires the application of pilots throughout an OFDM symbol, which adversely affects the bandwidth efficiency of the system. In addition, our simulations show that the approach in \cite{Rabiei-20010Nov} is outperformed by the receiver structure proposed in this paper.

\subsection{Contributions}
In this paper, a computationally efficient training based approach for joint channel, CFO, and PHN estimation in OFDM systems is presented. In order to detect the data symbols in the presence of time-varying PHN, an iterative receiver is proposed. The major contributions of this paper can be summarized as follows:\footnote{This paper was in part presented at IEEE International Workshop on Signal Processing Advances in Wireless Communications \cite{Omar-2013June}.}
\begin{itemize}
  \item We propose an expectation conditional maximization (ECM) based estimator for jointly obtaining the channel, CFO, and PHN parameters in OFDM systems. The ECM based estimation is carried out in two steps. In the expectation or E-step, an \emph{extended Kalman filter} (EKF) based estimator is utilized to accurately track the PHN over the training OFDM symbol. During the maximization or M-step, the channel and CFO parameters are estimated by minimizing the derived negative log likelihood function (cf. \eqref{eq-eqb5}).
  \item We derive an expression for the HCRB for the joint estimation of the channel, CFO and PHN in OFDM systems. Simulation results show that, compared to the existing algorithms in the literature, the mean square error (MSE) of the proposed algorithm is closer to the HCRB and the proposed algorithm outperforms the existing estimation algorithms at moderate-to-high signal-to-noise ratio (SNR).
 \item We propose a new iterative algorithm based on the EKF for data detection and tracking the unknown time-varying PHN throughout the OFDM data packet.
  \item We show that the proposed estimation and detection algorithms are computationally efficient, compared to existing algorithms in the literature. In addition, the proposed estimation and detection algorithms outperform existing algorithms in terms of both the uncoded and the coded bit error rate (BER) performance.
\end{itemize}
\subsection{Organiztion}
The remainder of the paper is organized as follows. Section \ref{sec-sec1} describes the system model and the assumptions used in this work. Section \ref{sec-sec2} derives the HCRB for joint channel, CFO and PHN estimation in OFDM systems. Section \ref{sec-sec3} describes the proposed ECM based estimator while Section \ref{sec-sec4} presents the proposed receiver for joint data detection and PHN tracking. Section \ref{sec-sec4} analyzes the complexity of the proposed estimation and data detection algorithms and compares it with existing schemes. Section \ref{sec-sec5} provides numerical and simulation results. Finally, Section \ref{sec-sec6} concludes the paper.
\subsection{Notations}
Superscripts $(\cdot)^\ast$, $(\cdot)^H$, and $(\cdot)^T$ denote the conjugate, the conjugate transpose, and the transpose operators, respectively. Bold face small letters, e.g., $\mathbf{x}$, are used for vectors, bold face capital alphabets, e.g., $\mathbf{X}$, are used for matrices, and $[\mathbf{X}]_{x,y}$ represents the entry in row \emph{x} and column \emph{y} of $\mathbf{X}$. $\mathbf{I}_X$, $\mathbf{0}_{X \times X}$, and $\mathbf{1}_{X \times X}$ denote the $X \times X$ identity, all zero, and all 1 matrices, respectively. The notation $\mathbf{X}(n_1:n_2,m_1:m_2)$ is used to denote a submatrix of $\mathbf{X}$ from row $n_1$ to row $n_2$ and from column $m_1$ to column $m_2$. $|\cdot|$ is the absolute value operator, $|\mathbf{x}|$ denotes the element-wise absolute value of a vector $\mathbf{x}$, and \text{diag}($\mathbf{x}$) is used to denote a diagonal matrix, where the diagonal elements are given by vector $\mathbf{x}$. $\mathbf{X} \succeq \bar{\mathbf{X}}$ indicates that matrix ($\mathbf{X} \succeq \bar{\mathbf{X}}$) is positive semi-definite. $\mathbb{E}_{x,y}[\cdot]$ denotes the expectation over \emph{x} and \emph{y}. $\Re\{\cdot\}$ and $\Im\{\cdot\}$ denote the real and imaginary parts of a complex quantity, respectively. $\nabla_{\mathbf{x}}$ and $\triangle_{\textbf{y}}^{\mathbf{x}}$ represent the first and the second-order partial derivatives operator, i.e., $\nabla_{\mathbf{x}} = [\frac{\partial}{\partial x_1},\cdots,\frac{\partial}{\partial x_N}]^T$ and $\triangle_{\mathbf{y}}^{\mathbf{x}} = \nabla_{\mathbf{y}} \times \nabla_{\mathbf{x}}^T$. $\mathcal{N}(\mu,\sigma^2)$ and $\mathcal{CN}(\mu,\sigma^2)$ denote real and complex Gaussian
distributions with mean $\mu$ and variance $\sigma^2$, respectively. $\otimes$ denotes circular convolution. Finally, $\dot{z}$ denotes the Jacobian of $z$.

\section{System Model} \label{sec-sec1}

\begin{figure*}[t]
  \begin{center}
    \includegraphics[scale=0.8]{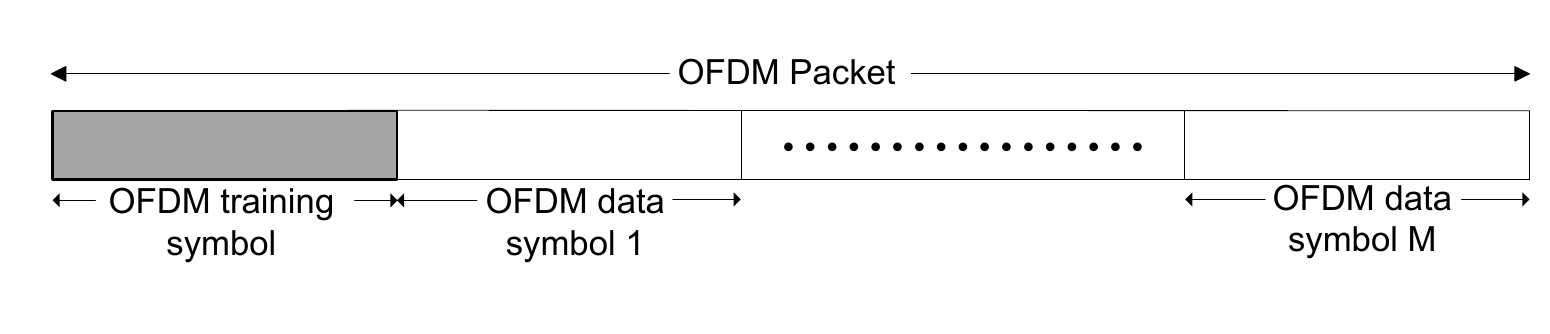}
   \end{center}
   \vspace{-35pt}
  \caption{Timing diagram for transmission of training and data symbols within an OFDM packet.}
  \label{fig-1}
\end{figure*}
We consider an OFDM packet of $(M+1)$ symbols, which consists of one training symbol and $M$ data symbols, as illustrated in Fig. \ref{fig-1}. In this paper, the following set of assumptions are adopted:
\begin{description}
  \item[A1.] The channel is modeled as a slow fading frequency-selective channel, i.e., the channel is assumed to be quasi-static, which is constant and \emph{unknown} over the OFDM packet duration and changes from packet to packet following a complex Gaussian distribution.
  \item[A2.] The time-varying PHN changes from symbol to symbol and is modeled as a Wiener process, i.e., $\theta_n=\theta_{n-1}+\delta_n$, $\forall$ $n$, where $\theta_n$ is the PHN at the $n$th instant, $\delta_n \sim \mathcal{N}(0,\sigma^2_\delta)$ is the PHN innovation and $ \sigma^2_\delta $ is the variance of the innovation process \cite{Demir-2000May},\cite{Lin-2006apr}.
  \item[A3.] The CFO is modeled as a \emph{deterministic unknown parameter} over a packet and is assumed to change from packet to packet.
  \item[A4.] The training symbol is assumed to be \emph{known} at the receiver.
  \item[A5.] The timing offset is assumed to be perfectly estimated. Hence, it is not considered in this paper.
\end{description}
Note that assumptions A1, A2, A3, and A5 are in line with previous channel, CFO and PHN estimation algorithms in \cite{{Armada-1998Jan},{Armada-2001Jun},{Lin-2006apr},{Lin-2006sept},{Lin-2007May},{Rabiei-20010Nov},{Carvajal-2013Jan}}. Assumption A2 is also reasonable in many practical scenarios to describe the behavior of practical oscillators \cite{Demir-2000May} \cite{Lin-2006sept}. Furthermore, assumption A4 is adopted in the IEEE $802.11$ac/ad standards to estimate channel and CFO in \cite{{Wu-2002Dec},{Munier-2008Jul},{Lin-2006sept},{Petrovic-2007Aug},{Rabiei-20010Nov},{Paul-2008FQ},{Zhu-2011May}}.

The complex baseband OFDM signal is given by
\begin{align} \label{eq-eq1}
x_n=\frac{1}{\sqrt{N}}\sum\limits^{N-1}_{k=0}d_{k}e^{{j2 \pi kn}/{N}} &&n=0,1,\ldots,N-1,
\end{align}
where $d_k$, for $k = 1,\hdots,N$, is the modulated training symbol, $x_n$ is the \emph{n}th sample of the transmitted OFDM symbol, \emph{N} is the number of subcarriers, and \emph{k} denotes the subcarrier index. At the receiver, after removing the cyclic prefix, the complex baseband received signal, $r_n$, is given by
\begin{subequations}\label{eq-eq2}
\begin{align}
r_n&=e^{j(\bar{\theta}_n+2 \pi n\epsilon/N)} \bar{s}_n+\eta_n \label{eq-eq2a} \\
&=e^{j(\theta_n+2 \pi n\epsilon/N)}s_n+\eta_n, \label{eq-eq2b}
\end{align}
\end{subequations}
where $\bar{s}_n \triangleq \bar{h}_n \otimes x_n$ is the received OFDM training symbol, $\{\bar{\theta}_n\}^{N-1}_{n=0}$ is the discrete-time PHN sequence, $\epsilon$ is the normalized CFO, $\{\bar{h}_l\}^{L-1}_{l=0}$ is the channel impulse response, \emph{L} is the channel length, and $\bar{h}_l \sim \mathcal{C} \mathcal{N}(\mu_{h_l},\sigma^2_{h_l})$. Note that \eqref{eq-eq2b} is an equivalent system model representation of \eqref{eq-eq2a}, where $s_n \triangleq h_n \otimes x_n$, $h_n \triangleq e^{j\bar{\theta}_0} \bar{h}_n$ and $ \theta_n \triangleq \bar{\theta}_n - \bar{\theta}_0$. This equivalent system model helps to distinguish between the phase disturbance caused by PHN and the channel phase for the first sample, which in turn resolves the phase ambiguity in the joint estimation problem as indicated in Section \ref{sec-sec3}. In addition, $\{\eta_n\}^{N-1}_{n=0}$ is the complex \emph{additive white Gaussian noise }(\emph{AWGN}) with zero-mean and known variance $\sigma^2_w$. The received signal, $\mathbf{r} \triangleq [r_0,r_1,\ldots,r_{N-1}]^T$, in vector form is given by
\begin{align} \label{eq-eq3}
\mathbf{r}=\mathbf{E}\mathbf{P}\mathbf{F}^H\mathbf{D}\mathbf{W}\mathbf{h}+\boldsymbol{\eta},
\end{align}
where
\begin{itemize}
\item $\mathbf{E} \triangleq \text{diag}([1,e^{(j 2 \pi \epsilon/N)},\ldots,e^{(j 2 \pi \epsilon/N) \times (N-1)}]^T)$ is the $N \times N$ CFO matrix,
\item $\mathbf{P} \triangleq \text{diag}([e^{j \theta_0}, e^{j \theta_1},\ldots,e^{j \theta_{N-1}}]^T)$ is the $N \times N$ PHN matrix,
\item $\mathbf{F}$ is an $N \times N$ DFT matrix, i.e.,  $ [\mathbf{F}]_{\emph{l},\emph{n}} \triangleq (1/ \sqrt{N}) e^{-j(2 \pi nl/N)}$ for $n,l=0,1,\cdots,N-1$,
\item $\mathbf{D} \triangleq \text{diag}(\mathbf{d}), \mathbf{d} \triangleq [d_0,d_1,\cdots,d_{N-1}]^T$ is the modulated training vector,
\item $\mathbf{W}$ is an $N \times L$ DFT matrix, i.e., $\mathbf{W} \triangleq \mathbf{F}(1:N,1:L)$,
\item \emph{L} denotes the number of channel taps,
\item $\mathbf{h} \triangleq [h_0,h_1,\ldots,h_{L-1}]^T$ is the channel impulse response (CIR), and
\item $\boldsymbol{\eta} \triangleq [\eta_0,\eta_1,\cdots,\eta_{N-1}]^T$ is the noise vector.
\end{itemize}
\section{Derivation of the hybrid Cram\'{e}r-Rao bound} \label{sec-sec2}
In this section, the HCRB for the joint estimation of the CIR, PHN, and CFO parameters in OFDM systems is derived. The HCRB is a lower bound on the joint estimation of random, e.g., PHN, and deterministic, e.g., CIR and CFO parameters. Let $\boldsymbol{\lambda}=[\boldsymbol{\theta}^T \ \Re\{\mathbf{h}\}^T \ \Im\{\mathbf{h}\}^T \ \epsilon]^T$ be the vector of hybrid parameters of interest, where $\boldsymbol{\theta} \triangleq [\theta_1,\ldots,\theta_{N-1}]^T$ is a vector of random PHN parameters and the channel vector, $\mathbf{h}$, and the CFO, $\epsilon$, are modeled as deterministic parameters. Note that it is clear from \eqref{eq-eq2b} that there is no need to estimate $\theta_0$.  The accuracy of estimating $\boldsymbol{\lambda}$ is lower bounded by the HCRB, $\boldsymbol{\Omega}$, as  \cite[pp. 1-85]{Trees-2007USA}
\begin{align} \label{eq:eqb4}
\mathbb{E}_{\mathbf{r},\boldsymbol{\theta}|\epsilon}\left[(\hat{\boldsymbol{\lambda}}(\mathbf{r})-\boldsymbol{\lambda})(\hat{\boldsymbol{\lambda}}(\mathbf{r})-\boldsymbol{\lambda})^T\right] \succeq \boldsymbol{\Omega}.
\end{align}
Let us define $\boldsymbol{\Omega}\triangleq\mathbf{B}^{-1}$. Here, $\mathbf{B}$ is an $(N+2L)\times(N+2L)$ hybrid information matrix (HIM), which is determined according to the following theorem.

\textbf{Theorem 1}: The closed-form HIM for joint estimation of CIR, PHN, and CFO is given by
\begin{align} \label{eq-eqb5HCRB}
\mathbf{B}=\frac{2}{\sigma^2_w}\Re\left\{\left[
             \begin{array}{cccc}
               \mathbf{B}_{11} & \mathbf{B}_{12} & \mathbf{B}_{13} & \mathbf{b}_{14} \\
               \mathbf{B}_{21} & \mathbf{B}_{22} & \mathbf{B}_{23} & \mathbf{b}_{24} \\
               \mathbf{B}_{31} & \mathbf{B}_{32} & \mathbf{B}_{33} & \mathbf{b}_{34} \\
               \mathbf{b}_{41} & \mathbf{b}_{42} & \mathbf{b}_{43} & \text{b}_{44} \\
             \end{array}
           \right]\right\},
\end{align}
where
\begin{itemize}
\item $\mathbf{B}_{11} \triangleq \bar{\mathbf{Q}}_1^H \bar{\mathbf{Q}}_1+\boldsymbol\Lambda$ is the $(N-1) \times (N-1)$ HIM for the estimation of $\boldsymbol{\theta}$,
             $\mathbf{Q}_1 = \text{diag}(\mathbf{F}^H\mathbf{D}\mathbf{W}\mathbf{h})$, and $\bar{\mathbf{Q}}_1 = \mathbf{Q}_1(2:N,2:N)$,
\item $\boldsymbol\Lambda$ is an $(N-1) \times (N-1)$ tridiagonal matrix with diagonal elements given by $\frac{\sigma^2_w}{2 \sigma^2_\delta} [ 1,2,\hdots,2,1  ]$ and off-diagonal elements given by $\frac{-\sigma^2_w}{2 \sigma^2_\delta} [ 1,\hdots,1  ]$,
\item $\mathbf{B}_{22} \triangleq  \mathbf{Q}_2^H \mathbf{Q}_2$ is an $L \times L$ information matrix for the estimation of real part of $\mathbf{h}$, and $\mathbf{Q}_2 = \mathbf{F}^H\mathbf{D}\mathbf{W}$,
\item $\mathbf{B}_{33} \triangleq  \mathbf{Q}_2^H \mathbf{Q}_2$ is an $L \times L$ information matrix for the estimation of imaginary part of $\mathbf{h}$,
\item $\text{b}_{44} \triangleq \mathbf{q}_5^H \mathbf{q}_5$ is a scalar representing the information for the estimation of CFO, $\epsilon$, $\mathbf{q}_5 = \sqrt{\mathbf{M}}\mathbf{F}^H\mathbf{D}\mathbf{W}\mathbf{h}$, and $\mathbf{M} \triangleq \text{diag} \left(\left[(2 \pi \frac{0}{N})^2, (2 \pi \frac{1}{N})^2,\ldots,(2 \pi \frac{N-1}{N})^2\right]^T\right)$,
\item $\mathbf{B}_{12} = \mathbf{B}_{21}^H \triangleq - j \bar{\mathbf{Q}}_1^H \bar{\mathbf{Q}}_2$, $\bar{\mathbf{Q}}_2 = \mathbf{Q}_2(2:N,1:L)$,
\item $\mathbf{B}_{13} = \mathbf{B}_{31}^H \triangleq \bar{\mathbf{Q}}_1^H \bar{\mathbf{Q}}_2$,
\item $\mathbf{b}_{14} = \mathbf{b}_{41}^H \triangleq \bar{\mathbf{Q}}_4^H \bar{\mathbf{q}}_3$, $\mathbf{Q}_4 = \text{diag}(\sqrt{\mathbf{M}}\mathbf{F}^H\mathbf{D}\mathbf{W}\mathbf{h})$, $\bar{\mathbf{Q}}_4 = \mathbf{Q}_4(2:N,2:N)$, $\mathbf{q}_3 = \mathbf{F}^H\mathbf{D}\mathbf{W}\mathbf{h}$, and $\bar{\mathbf{q}}_3 = \mathbf{q}_3(2:N)$,
\item $\mathbf{B}_{23} = \mathbf{B}_{32}^H \triangleq j \mathbf{Q}_2^H \mathbf{Q}_2$,
\item $\mathbf{b}_{24} = \mathbf{b}_{42}^H \triangleq j \mathbf{Q}_2^H \mathbf{q}_5$, and
\item $\mathbf{b}_{34} = \mathbf{b}_{43}^H \triangleq \mathbf{Q}_2^H \mathbf{q}_5$.
\end{itemize}

\emph{Proof:} The proof is given in Appendix A.

Finally, the HCRB, $\boldsymbol\Omega$, is given by the inverse of the HIM. i.e., $\boldsymbol\Omega = \mathbf{B}^{-1}$. Note that the HCRB of the channel, $\mathbf{h}$, is obtained by adding the HCRB for real and imaginary parts of channels.

\textit{Remark 1:} It is difficult to find the closed-form HCRB due to the complicated structure of the BIM. However, to obtain analytical insights, we consider a simple case, $N = 2$, $L =1$, and derive the closed-form HCRB for the joint estimation of CIR, PHN, and CFO. Due to space limitation, the details are omitted and the final results for the closed-form HCRB of the CIR and the CFO estimation are given below
\begin{subequations}\label{eq:cfHCRB}
\begin{align}\label{eq:cfHCRBb}
\text{HCRB}_{\text{CIR}}|_{N=2} = \frac{(2\gamma+\alpha)N^2 \sigma_w^2}{2(\alpha+\gamma) \gamma},
\end{align}
\begin{align}\label{eq:cfHCRBc}
\text{HCRB}_{\text{CFO}}|_{N=2} =  \frac{N^2 \left( (\alpha+\gamma) N^2 \sigma_w^2 + 2 \gamma \alpha |h|^2 \sigma^2_\delta  \right)}{ 8 \alpha \gamma |h|^2 \pi^2},
\end{align}
\end{subequations}
where $\alpha \triangleq |d_1 + d_2 e^{j \frac{2 \pi}{N} }|^2$ and $\gamma \triangleq |d_1 + d_2|^2$. It can be observed from \eqref{eq:cfHCRBb} that the HCRB for CIR estimation is independent of the PHN variance, $\sigma^2_\delta$. Moreover, according to \eqref{eq:cfHCRBc}, HCRB for CFO estimation is determined by PHN variance, $\sigma^2_\delta$, at high SNR. This interesting insight is also reflected through simulations results in Fig. \ref{fig-4} and Fig. \ref{fig-5}, respectively, in Section \ref{sec-sec5}.
\section{Proposed ECM Based Estimator} \label{sec-sec3}
In this section, an ECM based algorithm that utilizes the OFDM training symbol to jointly estimate the CIR, CFO and PHN at the receiver is derived. Joint data detection and PHN tracking during data transmission interval is analyzed in Section V.

\begin{figure}[t]
  \begin{center}
    \includegraphics[scale=1.0]{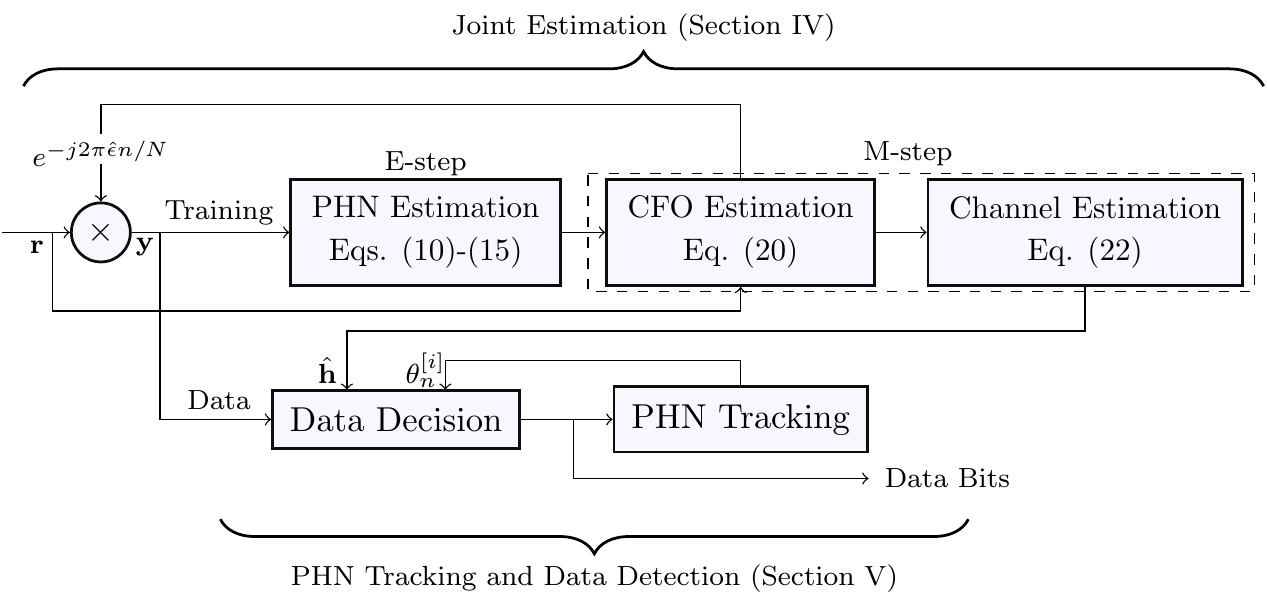}
   \end{center}
   \vspace{-0cm}
  \caption{Proposed estimator based on an ECM algorithm and data detection.}
  \label{fig-2}
\end{figure}

Fig. \ref{fig-2} depicts the block diagram of the overall system employing the proposed ECM based estimator and the joint data detection and PHN mitigation algorithm. As illustrated in Fig. \ref{fig-2}, the proposed ECM based estimator applies the training OFDM symbol at the beginning of each packet to estimate the CIR and CFO in the presence of PHN. Next, the data detection is performed by: 1) mitigating the impact of CFO over the length of packet by multiplying the received packet with the complex conjugate of the estimated CFO coefficients supplied by the proposed ECM estimator; 2) tracking the PHN parameters using an \emph{iterative} algorithm that utilizes an EKF; 3) mitigating the effect of PHN over the received packet; and 4) detecting data symbols using the estimated CIR and hard decision decoding.

As shown in Fig. \ref{fig-2}, the ECM algorithm iterates between the expectation step (E-step) and the maximization step (M-step). In the E-step, an EKF is used to update the PHN vector at the $(i+1)$th iteration, $\boldsymbol{\theta}^{[i+1]}$, using the CIR and CFO estimates, $\hat{\mathbf{h}}^{[i]}$ and $\hat{\epsilon}^{[i]}$, respectively, obtained from the previous iteration, i.e., $i$th iteration. Next, in the M-step, the estimates of the CIR and CFO at the ($i+1$)th iteration, $\hat{\mathbf{h}}^{[i+1]}$ and $\hat{\epsilon}^{[i+1]}$, respectively are obtained. 

For the given problem, the incomplete data set is given by the $N \times 1$ vector $\mathbf{s}\triangleq\mathbf{F}^H\mathbf{D}\mathbf{W}\mathbf{h}=[s_0,s_1,\ldots,s_{N-1}]^T$ and the received data, $\mathbf{r}$ in \eqref{eq-eq3}. Following \cite{Simon-2012Feb}, the hidden variable is chosen to be $\boldsymbol{\theta}$. Thus, the complete data set is defined as $\mathbf{z}\triangleq[\mathbf{r}^T \ \boldsymbol{\theta}^T]^T$. Moreover, the negative log likelihood function (LLF) of the complete data, $\log p(\mathbf{z};\epsilon, \mathbf{h})$, is given by
\begin{align} \label{eq-eqb5}
\log p(\mathbf{z};\epsilon)&= \text{C} +\frac{1}{\sigma^2_w}\sum_{n=0}^{N-1}\parallel r_n-e^{j 2 \pi \epsilon n/N}e^{j\theta_n}s_n \parallel ^2 + \log p(\theta_0)+\sum_{n=0}^{N-1}\log p(\theta_n|\theta_{n-1}),
\end{align}
where $\text{C}$ is a constant. Note that $s_n$ (defined below \eqref{eq-eq2b}) in \eqref{eq-eqb5} includes the CIR. The E and M-steps for estimating the CIR, CFO, and PHN in the training interval are detailed in the following subsections.

\subsection{E-step}\label{sec-sec3-estep}

In this step, the received signal $r_n$ is first multiplied by $e^{-j2 \pi \hat{\epsilon}^{[i]} n/N}$. Subsequently, the signal $y_n \triangleq e^{-j2 \pi n\epsilon^{[i]}/N}r_n$ is used to estimate the PHN vector, where $\hat{\epsilon}^{[i]}$ is the latest CFO estimate obtained from the previous iteration. We propose to use an EKF during the E-step to estimate the PHN samples $\boldsymbol{\theta}$. The intuition behind choosing the EKF will be explained shortly after (\ref{eq-eq6}).

The signal $y_n$ can be written as
\begin{equation} \label{eq-eq7}
y_n=e^{-j2 \pi n\epsilon^{[i]}/N}r_n=e^{j2 \pi n \Delta \hat{\epsilon}/N}e^{j\theta_n}s^{[i]}_n+\tilde{w}_n,
\end{equation}
where $s^{[i]}_n$ is the $n$th symbol of the vector $\mathbf{s}^{[i]} \triangleq\mathbf{F}^H\mathbf{D}\mathbf{W}\hat{\mathbf{h}}^{[i]}$, $\Delta \hat{\epsilon} \triangleq \epsilon-\hat{\epsilon}^{[i]}$, and $\tilde{w}_n \triangleq w_n e^{-j2 \pi n\hat{\epsilon}^{[i]}/N}$. The state and observation equations at time $n$ are given by
\begin{align} \label{eq-eq6}
\theta_n=&\theta_{n-1}+\delta_n,\\
\label{eq-eq7}
y_n=&z_n+w_n=e^{j\theta_n}s_n+\tilde{w}_n,
\end{align}
respectively. Since the observation equation in (\ref{eq-eq7}) is a non-linear function of the unknown state vector $\boldsymbol{\theta}$, the EKF is used instead of the simple Kalman filtering. The EKF uses the Taylor series expansion to linearize the non-linear observation equation in (\ref{eq-eq7}) about the current estimates \cite{Kay-1993}. Thus, the Jacobian of $z_n$, $\dot{z}_n$, is evaluated by computing the first order partial derivative of $z_n$ with respect to $\theta_n$ as
\begin{align} \label{eq-eq8}
\dot{z}_n=\frac{\partial z (\theta_n)}{\partial \theta_n}|_{\theta_n=\hat{\theta}_{n|n-1}} =& j z (\hat{\theta}_{n|n-1}) \\
=& j e^{j\hat{\theta}^{[i]}_{n|n-1}}\hat{s}_n.\nonumber
\end{align}
The first and second moments of the state vector at the $i$th iteration denoted by $\hat{\theta}^{[i]}_{n|n-1}$ and $M^{[i]}_{n|n-1}$, respectively, are given by
\begin{align} \label{eq-eq9}
\hat{\theta}^{[i]}_{n|n-1}=&\hat{\theta}^{[i]}_{n-1|n-1},
\\
 \label{eq-eq10}
M^{[i]}_{n|n-1}=&M^{[i]}_{n-1|n-1}+\sigma^2_\delta.
\end{align}
Given the observation $y_n$, the Kalman gain $K_n$, posteriori state estimate $\hat{\theta}^{[i]}_{n|n}$, and the filtering error covariance, $M^{[i]}_{n|n}$ are given by
\begin{align} \label{eq-eq11}
K_n=&M^{[i]}_{n|n-1}\dot{z}^\ast(\theta_{n|n-1})\big(\dot{z}(\theta_{n|n-1})M^{[i]}_{n-1|n-1}  \dot{z}^\ast(\theta_{n|n-1})+\sigma^2_w\big)^{-1},
\\
 \label{eq-eq12}
\hat{\theta}^{[i]}_{n|n}=&\hat{\theta}^{[i]}_{n|n-1}+\Re\big\{K_n\big(y_n-e^{j\hat{\theta}^{[i]}_{n|n-1}}\hat{s}_n^{[i]}\big)\big\},
\\
 \label{eq-eq13}
M^{[i]}_{n|n}=&\Re\big\{M^{[i]}_{n|n-1}-K_n\dot{z}(\theta_{n|n-1})M^{[i]}_{n|n-1}\big\},
\end{align}
respectively. Before starting the EKF recursion (\ref{eq-eq8})-(\ref{eq-eq13}), $\hat{\theta}^{[0]}_{1|0}$ and $M^{[0]}_{1|0}$ are initialized to $\hat{\theta}^{[0]}_{1|0}=0$ and $M^{[0]}_{1|0}=\sigma^2_\delta$. The initialization choice for the PHN follows from the assumption that the complex channel parameter takes into account the PHN corresponding to the first symbol.

\subsection{M-step}
In this step, the CIR and CFO are estimated by minimizing the LLF in (\ref{eq-eqb5}). In order to further reduce the complexity associated with the \emph{M}-step, the minimization in (\ref{eq-eqb5}) is carried out with respect to one of the parameters while keeping the remaining parameters at their most recently updated values \cite{Meng-1993Meng}, \cite{McLachlan-2008Wiley}. First, by using the channel estimate at the $i$th iteration, $\hat{\mathbf{h}}^{[i]}$, and the PHN vector estimate from the E-step, $\hat{\boldsymbol{\theta}}^{[i+1]}$, the LLF in (\ref{eq-eqb5}) is minimized with respect to $\epsilon$ to obtain the CFO estimate for the $(i+1)$th iteration, $\hat{\epsilon}^{[i+1]}$, as
\begin{equation} \label{eq-eq122}
\hat{\epsilon}^{[i+1]}= \arg \underset{\epsilon}{\min} \sum_{n=0}^{N-1} \parallel r_n-e^{j 2 \pi \epsilon n/N}e^{j\theta_n}s_n \parallel ^2\big|_{{\theta_n=\hat{\theta}_n^{[i]}},\mathbf{h}=\hat{\mathbf{h}}^{[i]}}.
\end{equation}
After simplifying (\ref{eq-eq122}), we have
\begin{equation} \label{eq-eq2}
\hat{\epsilon}^{[i+1]}= \arg \underset{\epsilon}{\max} \sum_{n=0}^{N-1} \Re\{(r_n)^{\ast} \hat{S}_n^{[i]}e^{j 2 \pi \epsilon n/N}\},
\end{equation}
where $\hat{S}_n^{[i]}= e^{j\hat{\theta}^{[i]}_n}\hat{s}_n$. In order to resolve the nonlinearity in (\ref{eq-eq2}), we can approximate the term $e^{j 2 \pi \epsilon n/N}$ using a second order Taylor series expansion around the pervious CFO estimate, $\hat{\epsilon}^{[i]}$, as
\begin{align} \label{eq-eq17}
e^{j 2 \pi \epsilon n/N}=&e^{j 2 \pi \hat{\epsilon}^{[i]} n/N}+(\epsilon-\hat{\epsilon}^{[i]})(j\frac{2\pi}{N}n)e^{j 2 \pi \hat{\epsilon}^{[i]} n/N} + \frac{1}{2}(\epsilon-\hat{\epsilon}^{[i]})^2(j\frac{2\pi}{N}n)^2e^{j 2 \pi \hat{\epsilon}^{[i]} n/N}.
\end{align}
Substituting \eqref{eq-eq17} into \eqref{eq-eq2}, $\hat{\epsilon}^{[i+1]}$ is given by
\begin{align} \label{eq-eq18}
\hat{\epsilon}^{[i+1]}=& \arg \underset{\epsilon}{\max} \bigg\{\sum_{n=0}^{N-1} \Re \big\{(r_n)^{\ast} \hat{S}_n^{[i]} e^{j 2 \pi \hat{\epsilon}^{[i]} n/N} \\
&+ (\epsilon-\hat{\epsilon}^{[i]}) \sum_{n=0}^{N-1} \Re \big\{(r_n)^{\ast} \hat{S}_n^{[i]} (j\frac{2\pi}{N}n) e^{j 2 \pi \hat{\epsilon}^{[i]} n/N}\big\} \nonumber\\
&+\frac{1}{2} (\epsilon-\hat{\epsilon}^{[i]})^2 \sum_{n=0}^{N-1} \Re \big\{(r_n)^{\ast} \hat{S}_n^{[i]} (j\frac{2\pi}{N}n)^2 e^{j 2 \pi \hat{\epsilon}^{[i]} n/N}\big\}\bigg\}\nonumber.
\end{align}
Taking the derivative of (\ref{eq-eq18}) with respect to $\epsilon$ and equating the result to zero, the estimate of $\epsilon$ at the $(i+1)$th iteration is given by
\begin{equation} \label{eq-eq191}
\hat{\epsilon}^{[i+1]}=\hat{\epsilon}^{[i]}+\frac{N}{2 \pi} \frac{\sum_{n=0}^{N-1}n\Im\big\{(r_n)^{\ast} \hat{S}_n^{[i]} e^{j 2 \pi \hat{\epsilon}^{[i]} n/N}\big\}}{\sum_{n=0}^{N-1} n^2\Re\big\{(r_n)^{\ast} \hat{S}_n^{[i]} e^{j 2 \pi \hat{\epsilon}^{[i]} n/N}\big\}}.
\end{equation}

Next, by setting $\boldsymbol{\theta}$ and $\epsilon$ to their latest updated values, the updated value of $\hat{\mathbf{h}}$ at the $(i+1)$th iteration, $\hat{\mathbf{h}}^{[i+1]}$, is determined as outlined below.

Based on the vectorial form of the received signal in ($\ref{eq-eq3}$), the negative LLF, in (\ref{eq-eqb5}), can be written as
\vspace{-10pt}
\begin{align} \label{eq-eqb41}
\log p(\mathbf{z};\epsilon)= \text{C}_1 +\parallel \mathbf{r}-\mathbf{E}\mathbf{P}\boldsymbol{\Gamma}\mathbf{h} \parallel ^2+\log p(\boldsymbol{\theta}).
\end{align}
where $\boldsymbol{\Gamma} \triangleq \mathbf{F}^H\mathbf{D}\mathbf{W}$ and $\text{C}_1$ is a constant. Taking the derivative of (\ref{eq-eqb41}) with respect to $\mathbf{h}$ and equating the result to zero, the estimate of $\mathbf{h}$ at the $(i+1)$th iteration is given by
\begin{eqnarray} \label{eq-eq42}
\hat{\mathbf{h}}^{[i+1]}&=&(\boldsymbol{\Gamma}^H \boldsymbol{\Gamma})^{-1}\boldsymbol{\Gamma}^H\hat{\mathbf{P}}^H\hat{\mathbf{E}}^H\mathbf{r},
\end{eqnarray}
where $\hat{\mathbf{E}} \triangleq \text{diag}([e^{(j 2 \pi \hat{\epsilon}^{[i+1]}/N)\times 0},e^{(j 2 \pi \hat{\epsilon}^{[i+1]}/N)}, \ldots  ,e^{(j 2 \pi \hat{\epsilon}^{[i+1]}/N) \times (N-1)}]^T)$ and $\hat{\mathbf{P}} \triangleq \text{diag}([e^{j \hat{\theta}^{[i]}_0}, e^{j \hat{\theta}^{[i]}_1},\ldots,e^{j \hat{\theta}^{[i]}_{N-1}}]^T)$. Note that $\hat{\epsilon}^{[i+1]}$ and $\hat{\boldsymbol{\theta}}^{[i]} \triangleq [\hat{\theta}^{[i]}_1,\hat{\theta}^{[i]}_2,\ldots,\hat{\theta}^{[i]}_{N-1}]^T$ are determined as in (\ref{eq-eq191}) and (\ref{eq-eq12}), respectively.

Using (\ref{eq-eq12}), (\ref{eq-eq191}), and (\ref{eq-eq42}), the proposed algorithm iteratively updates the PHN, CFO, and CIR estimates, respectively. The algorithm is terminated when the difference between the likelihood functions of two iterations is smaller than a threshold $\zeta$, i.e.,
\begin{align} \label{eq-eq43}
\Bigg |\sum_{n=0}^{N-1}& \left\| r_n-e^{j 2 \pi \hat{\epsilon}^{[i+1]} n/N}e^{j \hat{\theta}_n^{[i+1]}}s_n^{[i+1]} \right\| ^2 \nonumber\\
&- \sum_{n=0}^{N-1}\left\| r_n-e^{j 2 \pi \hat{\epsilon}^{[i]} n/N}e^{j \hat{\theta}_n^{[i]}}s_n^{[i]} \right\| ^2 \Bigg | \leq \zeta.
\end{align}

\subsection{Initialization and Convergence}
The appropriate initialization of CFO and CIR, i.e., $\hat{\epsilon}^{[0]}$ and $\hat{\mathbf{h}}^{[0]}$, respectively, is essential to ensure the global convergence of the proposed estimator \cite{Fesler-94-A}. The initialization process can be summarized as follows:
\begin{itemize}
\item The initial CFO estimate is obtained by applying an exhaustive search for the value of $\epsilon$ that minimizes the cost function, $\sum_{n=0}^{N-1}\parallel r_n-e^{j 2 \pi \hat{\epsilon} n/N}\hat{s}_n \parallel ^2$. Here, $\hat{s}_n$ is the \emph{n}th symbol of the vector $\hat{\mathbf{s}} \triangleq \mathbf{F}^H\mathbf{D}\mathbf{W}\hat{\mathbf{h}}$ with $\hat{\mathbf{h}} \triangleq  (\boldsymbol{\Gamma}^H \boldsymbol{\Gamma})^{-1}\boldsymbol{\Gamma}^H\hat{\mathbf{E}}^H\mathbf{r}$. Note that this exhaustive search needs to be only carried out at the system start up to initialize the estimation process. Simulations in Section \ref{sec-sec5} indicate that an exhaustive search with a coarse step size of $10^{-2}$ is sufficient for the initialization of the proposed estimator.

\item Using $\hat{\epsilon}^{[0]}$, the initial channel estimate, $\hat{\mathbf{h}}^{[0]}$, is obtained by applying the relationship, $\hat{\mathbf{h}}^{[0]}= (\boldsymbol{\Gamma}^H \boldsymbol{\Gamma})^{-1}\boldsymbol{\Gamma}^H \\ (\hat{\mathbf{E}}^{[0]})^H\mathbf{r}$. Here, $\hat{\mathbf{E}}^{[0]} = \hat{\mathbf{E}}|_{\hat{\epsilon} = \hat{\epsilon}^{[0]}}$.
\end{itemize}
Note that based on the equivalent system model in \eqref{eq-eq2b} and the simulation results in Section \ref{sec-sec5}, it can be concluded that the proposed ECM algorithm converges globally when the PHN vector $\hat{\boldsymbol{\theta}}$ is initialized as $\hat{\boldsymbol{\theta}}^{[0]} = \mathbf{0}_{N \times 1}$.

Simulation results in Section \ref{sec-sec5} show that at SNRs of $20$ dB or higher the proposed ECM-based estimator always converges to the true estimates in only $2$ iterations.

\section{Joint Data Detection and PHN Mitigation} \label{sec-sec4}
In this section, we propose an iterative detector that utilizes an EKF to the track the PHN parameters during the data transmission interval.

At first, using the estimated CFO value, the effect of CFO on the received data symbol, $\mathbf{r}$, in \eqref{eq-eq3} is compensated. As shown Fig. \ref{fig-2}, the resulting signal, $\mathbf{y} \triangleq [y_1,\hdots,y_n]$, where $y_n$ is defined in \eqref{eq-eq7}, passes through an \emph{iterative} data detection and PHN estimation block. We propose to use an EKF to track the PHN samples,  $\boldsymbol{\theta}$, over the data symbols. The PHN estimation is similar to that in (\ref{eq-eq8})-(\ref{eq-eq13}) and is not presented here to avoid repetition. However, instead of training-based PHN tracking, the PHN estimation is followed in decision-directed fashion for the received data symbols. In other words, the estimate of the data symbol in the previous iteration, $\hat{\mathbf{d}}^{[i-1]}$, is used to update the symbol's PHN estimate at the current iteration $\hat{\boldsymbol{\theta}}^{[i]}$. Particularly, $\mathbf{s}^{[i]}$ in (\ref{eq-eq7}), is calculated as $\mathbf{s}^{[i]} = \mathbf{F}^H\hat{\mathbf{D}}^{[i-1]}\mathbf{W}\hat{\mathbf{h}}$, where $\hat{\mathbf{h}}$ is the CIR vector estimate obtained from the ECM estimator during the training interval, and $\hat{\mathbf{D}}^{[i-1]} \triangleq \text{diag}(\hat{\mathbf{d}}^{[i-1]})$.
Next, the data vector estimate is updated for the $i$th iteration. Following \cite{Lin-2007May} and based on the received signal in ($\ref{eq-eq3}$), the negative LLF for the CFO compensated signal, $\mathbf{y}$, can be written as
\begin{align} \label{eq-eqb53}
\log p(\mathbf{y},\hat{\mathbf{d}},\hat{\boldsymbol{\theta}})= \text{C} +\frac{1}{2\sigma_w^2}\parallel \mathbf{y}-\hat{\mathbf{P}}\mathbf{F}^H\hat{\boldsymbol{\Upsilon}}\hat{\mathbf{d}} \parallel ^2+\frac{1}{2\xi_d}\parallel \hat{\mathbf{d}} \parallel ^2+\log p(\boldsymbol{\theta}),
\end{align}
where C is a constant and
\begin{itemize}
  \item $\hat{\boldsymbol{\Upsilon}} \triangleq \text{diag}(\mathbf{W}\hat{\mathbf{h}})$ is the estimated channel frequency response,
  \item $\hat{\mathbf{d}} \triangleq [\hat{d}_0,\hat{d}_1,\cdots,\hat{d}_{N-1}]^T$ is the estimate of the modulated data vector, and
  \item $\xi_d$ is the average transmitted symbol power and normalized to 1,
\end{itemize}
Taking the derivative of (\ref{eq-eqb53}) with respect to $\mathbf{d}$ and equating the result to zero, the estimate of $\mathbf{d}$ at the $i$th iteration, $\hat{\mathbf{d}}^{[i]}$ is given by
\begin{align} \label{eq-eq54}
\hat{\mathbf{d}}^{[i]}=(\hat{\boldsymbol{\Upsilon}}^H\hat{\boldsymbol{\Upsilon}}+\frac{\sigma_w^2}{\xi_d}\mathbf{I}_N)^{-1}\hat{\boldsymbol{\Upsilon
}}^H\mathbf{F}\hat{\mathbf{P}}^H\mathbf{y},
\end{align}
where $\hat{\mathbf{P}} \triangleq \text{diag}([e^{j \hat{\theta}^{[i]}_0}, e^{j \hat{\theta}^{[i]}_1},\ldots,e^{j \hat{\theta}^{[i]}_{N-1}}]^T)$ and $\hat{\boldsymbol{\theta}}^{[i]} \triangleq [\hat{\theta}^{[i]}_1,\hat{\theta}^{[i]}_2,\ldots,\hat{\theta}^{[i]}_{N-1}]^T$ are obtained via the EKF based estimator.

Using the EKF set of equations \eqref{eq-eq8}$-$\eqref{eq-eq13} and (\ref{eq-eq54}), the proposed algorithm iteratively updates the PHN and data estimates, respectively, and stops when the difference between likelihood functions of two iterations is smaller than a threshold $\zeta$, i.e.,
\begin{align} \label{eq-eq55}
\Bigg |\sum_{n=0}^{N-1}& \left\| y_n-e^{j \hat{\theta}_n^{[i+1]}}\hat{s}_n^{[i+1]} \right\| ^2 - \sum_{n=0}^{N-1}\left\| y_n-e^{j \hat{\theta}_n^{[i]}}\hat{s}_n^{[i]} \right\| ^2 \Bigg | \leq \zeta.
\end{align}

Let $\hat{\mathbf{d}}^{[0]}$ denote the initial estimate of the transmitted data vector. Appropriate initialization of $\hat{\mathbf{d}}^{[0]}$ results in the proposed iterative detector to converge quickly. In our algorithm, the initial data estimate is obtained using $\hat{\mathbf{d}}^{[0]}=(\hat{\boldsymbol{\Upsilon}}^H\hat{\boldsymbol{\Upsilon}}+\frac{\sigma_w^2}{\xi_d}\mathbf{I}_N)^{-1}
\hat{\boldsymbol{\Upsilon}}^H\mathbf{F} \hat{\mathbf{P}}^H_{[m-1]}\mathbf{y}$, where $\hat{\mathbf{P}}_{[m-1]}$ is the PHN matrix estimate obtained from the previous OFDM symbol. Simulation results in Section~\ref{sec-sec5} indicate that at SNR$=20$ dB the proposed detector, on average, converges after $2$ iterations. The overall estimation and detection algorithm is summarized in Algorithm 1 on the next page.

\begin{algorithm} \label{eqn-JSCCalgorithm}
\caption{PROPOSED ECM ESTIMATOR AND DATA DETECTION ALGORITHMS}
\begin{algorithmic}
\STATE \textbf{ECM ESTIMATOR}
\STATE \underline{Initialization}
\STATE $\hat{\theta}^{[0]}_{1|0}=0$ and $M^{[0]}_{1|0}=\sigma^2_\delta$ and obtain $\hat{\epsilon}^{[0]}$ and $\hat{\mathbf{h}}^{[0]}$ using an exhaustive search and (\ref{eq-eq42}) with coarse step size i.e., $10^{-2}$
\WHILE {$\Bigg |\sum_{n=0}^{N-1} \left\| r_n-e^{j 2 \pi \hat{\epsilon}^{[i+1]} n/N}e^{j \hat{\theta}_n^{[i+1]}}s_n^{[i+1]} \right\| ^2 - \sum_{n=0}^{N-1}\left\| r_n-e^{j 2 \pi \hat{\epsilon}^{[i]} n/N}
e^{j \hat{\theta}_n^{[i]}}s_n^{[i]} \right\| ^2 \Bigg | > \zeta.$}
\FOR {$n=0,1,\ldots,N-1$}
\STATE {\eqref{eq-eq8}$-$\eqref{eq-eq13}}
\ENDFOR
\FOR {$n=0,1,\ldots,N-1$}
\STATE {$\hat{\epsilon}^{[i+1]}=\hat{\epsilon}^{[i]}+\frac{N}{2 \pi} \frac{\sum_{n=0}^{N-1}n\Im\big\{(r_n)^{\ast} \hat{S}_n^{[i]} e^{j 2 \pi \hat{\epsilon}^{[i]} n/N}\big\}}{\sum_{n=0}^{N-1} n^2\Re\big\{(r_n)^{\ast} \hat{S}_n^{[i]} e^{j 2 \pi \hat{\epsilon}^{[i]} n/N}\big\}}$}
\ENDFOR
\STATE {$\hat{\mathbf{h}}^{[i+1]}=(\boldsymbol{\Gamma}^H \boldsymbol{\Gamma})^{-1}\boldsymbol{\Gamma}^H\hat{\mathbf{P}}^H\hat{\mathbf{E}}^H\mathbf{r}$}
\STATE {$\hat{\mathbf{h}}^{[i]}=\hat{\mathbf{h}}^{[i+1]}$, $\hat{\boldsymbol{\theta}}^{[i]}=\hat{\boldsymbol{\theta}}^{[i+1]}$, $\hat{\epsilon}^{[i]}=\hat{\epsilon}^{[i+1]}$}
\ENDWHILE
\STATE \textbf{DATA DETECTION}
\FOR {$m=1,\hdots,M$}
\STATE \underline{Initialization}
\STATE Obtain $\hat{\mathbf{d}}^{[0]}=(\hat{\boldsymbol{\Upsilon}}^H\hat{\boldsymbol{\Upsilon}}+\frac{\sigma_w^2}{\xi_d}\mathbf{I}_N)^{-1}
\hat{\boldsymbol{\Upsilon}}^H\mathbf{F} \hat{\mathbf{P}}^H_{[m-1]}\mathbf{y}$
\STATE Replace $\hat{\mathbf{d}}^{[0]}$ by its hard decision.
\WHILE {$\Bigg |\sum_{n=0}^{N-1} \left\| y_n-e^{j \hat{\theta}_n^{[i+1]}}\hat{s}_n^{[i+1]} \right\| ^2 - \sum_{n=0}^{N-1}\left\| y_n-e^{j \hat{\theta}_n^{[i]}}\hat{s}_n^{[i]} \right\| ^2 \Bigg | > \zeta$}
\STATE Using the EKF set of equation in Section \ref{sec-sec3-estep} to estimate the PHN parameters,
\STATE {$\hat{\mathbf{d}}^{[i]}=(\hat{\boldsymbol{\Upsilon}}^H\hat{\boldsymbol{\Upsilon}}+\frac{\sigma_w^2}{\xi_d}\mathbf{I}_N)^{-1}\hat{\boldsymbol{\Upsilon}}^H\mathbf{F}\hat{\mathbf{P}}^H\mathbf{y}$}.
\STATE Replace $\hat{\mathbf{d}}^{[i]}$ by its hard decision.
\STATE {$\hat{\mathbf{d}}^{[i]}=\hat{\mathbf{d}}^{[i+1]}$}
\ENDWHILE
\ENDFOR
\end{algorithmic}
\end{algorithm}

\subsection{Complexity Analysis} \label{sec-CA}
In this subsection, the computational complexity of the proposed estimator and detector is compared with that of \cite{Lin-2006sept} and \cite{Lin-2007May}. Throughout this section, computational complexity is defined as the number of complex additions and multiplications \cite{Hani-2011Apr}.

Let us denote the computational complexity of the proposed estimator by $C_\text{EST}=C_\text{EST}^{[M]}+C_\text{EST}^{[A]}$, where $C_\text{EST}^{[M]}$ and $C_\text{EST}^{[A]}$ denote the number of complex multiplications and additions used by the estimator, respectively. $C_\text{EST}^{[M]}$ and $C_\text{EST}^{[A]}$ are determined as
\begin{align} \label{eq-eq59}
C_\text{EST}^{[M]}=&\big[\underbrace{N}_{(\ref{eq-eq8})}+\underbrace{5N}_{(\ref{eq-eq11})}+\underbrace{2N}_{(\ref{eq-eq12})}+\underbrace{2N}_{(\ref{eq-eq13})}+\underbrace{7N}_{(\ref{eq-eq191})}+\underbrace{LN(2N+1)}_{(\ref{eq-eq42})}
+\underbrace{N(N^2+L(N+1))}_{{s}_n \text{in} (\ref{eq-eq12})}\big]t_\text{ECM}\nonumber\\
&+\big[\underbrace{3N}_{\sum_{n=0}^{N-1}\parallel r_n-e^{j 2 \pi \hat{\epsilon} n/N}\hat{s}_n \parallel ^2}+\underbrace{LN(2N+1)}_{\hat{\mathbf{h}} \triangleq  {\xi_d}^{-1}\boldsymbol{\Gamma}^H\hat{\mathbf{E}}^H\mathbf{r}}
+\underbrace{N(N^2+L(N+1))}_{\hat{\mathbf{s}} \triangleq \mathbf{F}^H\mathbf{D}\mathbf{W}\hat{\mathbf{h}}}\big]t_\text{initialize}\nonumber\\
&+\underbrace{N^2(N+L)}_{\boldsymbol{\Gamma} \text{in} (\ref{eq-eq42})},
\end{align}
\begin{align} \label{eq-eq60}
C_\text{EST}^{[A]}=&\big[\underbrace{N}_{(\ref{eq-eq10})}+\underbrace{N}_{(\ref{eq-eq11})}+\underbrace{2N}_
{(\ref{eq-eq12})}+\underbrace{N}_{(\ref{eq-eq13})}+\underbrace{2N+1}_{(\ref{eq-eq191})}+\underbrace{L(N-1)(2N+1)}_{(\ref{eq-eq42})}
\nonumber\\
&+\underbrace{N(N-1)(L+1)+N(L-1)}_{{s}_n \text{in} (\ref{eq-eq12})}\big]t_\text{ECM} \nonumber\\
&+\big[\underbrace{2N}_{\sum_{n=0}^{N-1}\parallel r_n-e^{j 2 \pi \hat{\epsilon} n/N}\hat{s}_n \parallel ^2}+\underbrace{L(N-1)(2N+1)}_{\hat{\mathbf{h}} \triangleq  {\xi_d}^{-1}\boldsymbol{\Gamma}^H\hat{\mathbf{E}}^H\mathbf{r}}
+\underbrace{N(N-1)(L+1)+N(L-1)}_{\hat{\mathbf{s}} \triangleq \mathbf{F}^H\mathbf{D}\mathbf{W}\hat{\mathbf{h}}}\big]t_\text{initialize}\nonumber\\
&+\underbrace{N(N-1)(N+L)}_{\boldsymbol{\Gamma} \text{in} (\ref{eq-eq42})},
\end{align}
where $t_\text{ECM}$ is the number of iterations in the ECM estimator and $t_\text{initialize}$ is the number of iterations required to initialize the ECM algorithm. The latter depends on the step size of the exhaustive search used to initialize the CFO estimates.

Similarly, the computational complexity of the proposed detector is denoted by $C_\text{DATA DET}^{[M]}$ and $C_\text{DATA DET}^{[A]}$, where $C_\text{DATA DET}^{[M]}$ and $C_\text{DATA DET}^{[A]}$ denote the number of complex multiplications and additions used by the estimator, respectively. $C_\text{DATA DET}^{[M]}$ and $C_\text{DATA DET}^{[A]}$ are determined as
\begin{align} \label{eq-eq61}
C_\text{DATA DET}^{[M]}=&\big[\underbrace{N}_{(\ref{eq-eq8})}+\underbrace{5N}_{(\ref{eq-eq11})}+\underbrace{2N}_{(\ref{eq-eq12})}+\underbrace{2N}_{(\ref{eq-eq13})}+\underbrace{N(N^2+L(N+1))}_{{s}_n \text{in} (\ref{eq-eq12})}+\underbrace{N^2(5N+1)}_{(\ref{eq-eq54})}\big]t_\text{DATA DET} \nonumber\\
&+\underbrace{N^2(5N+1)}_{\hat{\mathbf{d}}^{[0]}=(\hat{\boldsymbol{\Upsilon}}^H\hat{\boldsymbol{\Upsilon}}+\frac{\sigma_w^2}{\xi_d}\mathbf{I}_N)^{-1}\hat{\boldsymbol{\Upsilon}}^H\mathbf{F}\mathbf{P}^H_{[m-1]}\mathbf{y}} +\underbrace{NL}_{\hat{\boldsymbol{\Upsilon}} \text{in} (\ref{eq-eq54})},
\end{align}
\begin{align} \label{eq-eq62}
C_\text{DATA DET}^{[A]}=&\big[\underbrace{N}_{(\ref{eq-eq10})}+\underbrace{N}_{(\ref{eq-eq11})}+\underbrace{2N}_{(\ref{eq-eq12})}
+\underbrace{N}_{(\ref{eq-eq13})}+\underbrace{N(N-1)(L+1)+N(L-1)}_{{s}_n \text{in} (\ref{eq-eq12})} \nonumber\\
&+\underbrace{N(N^2+N(N-1)(4N+1))}_{(\ref{eq-eq54})}\big]t_\text{DATA DET} +\underbrace{N(N^2+N(N-1)(4N+1))}_{\hat{\mathbf{d}}^{[0]}=(\hat{\boldsymbol{\Upsilon}}^H\hat{\boldsymbol{\Upsilon}}+\frac{\sigma_w^2}
{\xi_d}\mathbf{I}_N)^{-1}\hat{\boldsymbol{\Upsilon}}^H\mathbf{F}\mathbf{P}^H_{[m-1]}\mathbf{y}} \nonumber\\
&+\underbrace{N(L-1)}_{\hat{\boldsymbol{\Upsilon}} \text{in} (\ref{eq-eq54})},
\end{align}
where $t_\text{DATA DET}$ is the number of iterations required by the detector in \eqref{eq-eq55}.

Following similar steps as in (\ref{eq-eq59})-(\ref{eq-eq62}), we can find the computational complexity of the estimator in \cite{Lin-2006sept} as
\begin{align} \label{eq-eq63}
C_\text{[8,EST]}^{[M]}=\big[N^2(11N+7)+2N\big]t_{[8]}+N^2(9N+4L+1)+LN
\end{align}
\begin{align} \label{eq-eq64}
C_\text{[8,EST]}^{[A]}=\big[2N^3+(N-1)(9N^2+7N+2)+1\big]t_{[8]}+2N^3+(N-1)(N(7N+4L+1)+L)
\end{align}
where $t_{[8]}$ is the number of iterations required for estimating the CFO via an exhaustive search in \cite{Lin-2006sept}. Moreover, the notations $C_\text{[8,EST]}^{[M]}$ and $C_\text{[8,EST]}^{[A]}$ are used to denote the number of complex multiplications and additions used by the estimator in \cite{Lin-2006sept}, respectively.

The computational complexity of the detector in \cite{Lin-2007May} is given by
\begin{align} \label{eq-eq65}
C_\text{[20, DATA DET]}^{[M]}=\big[N^2(11N+6)\big]t_{[20]}+N^2(6N+1)
\end{align}
\begin{align} \label{eq-eq66}
C_\text{[20, DATA DET]}^{[A]}=\big[N(N-1)(9N+6)+N^2(2N+1)\big]t_{[20]}+N^2(6N-5)+N(N-1)
\end{align}
where $t_{[20]}$ is the number of iterations used by the detector in \cite{Lin-2007May}. Note that since the estimation approach of \cite{Lin-2006sept} and \cite{Tao-2009Oct} are similar, the computational complexity of the estimation algorithm in \cite{Tao-2009Oct} can be calculated using \eqref{eq-eq63} and \eqref{eq-eq64}. Note that we do not present the computational complexity of the algorithm in \cite{Rabiei-20010Nov} since the approach in \cite{Rabiei-20010Nov} only considers channel and PHN estimation while assuming that no CFO is present.

\begin{figure}[h!] \label{fig:complexourECMEKFversus820}
  \begin{center}
    \includegraphics[scale=0.6]{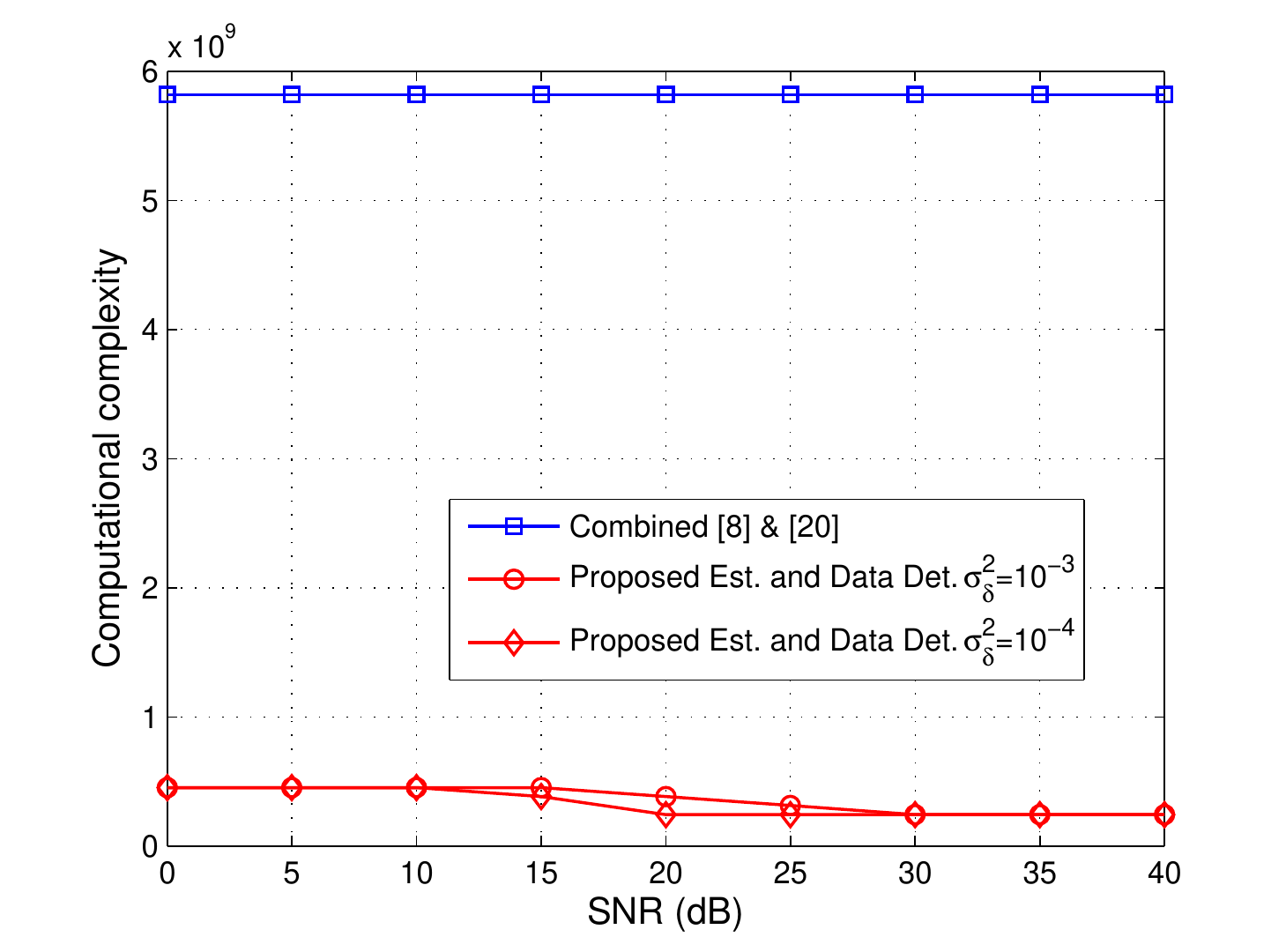}
   \end{center}
   \caption{Comparison of the computational complexity of the proposed algorithms and the algorithms in \cite{Lin-2006sept} and \cite{Lin-2007May} for PHN variance, $\sigma^2_\delta= [10^{-3}, 10^{-4}] \ \text{rad}^2$.}
\end{figure}

Fig. 3 compares the computational complexity of the proposed algorithm ((27)-(30)) and the existing algorithms in \cite{Lin-2006sept} and \cite{Lin-2007May} ((31)-(34)) for PHN variance, $\sigma^2_\delta= [10^{-3}, 10^{-4}] \ \text{rad}^2$. Note that the combined algorithm, \cite{Lin-2006sept} \& \cite{Lin-2007May}, is used because the authors (Lin et. al.) have proposed the estimation algorithm in \cite{Lin-2006sept} and the data detection algorithm in \cite{Lin-2007May}. For the comparison, the number of iterations, $t_\text{ECM}$, $t_\text{DATA DET}$ and $t_\text{initialize}$ for the proposed algorithm and $t_{[8]}$ and $t_{[20]}$ for the existing algorithms are determined as follows. \emph{For the proposed algorithm}, simulations indicate that (i) at low SNR, i.e., $SNR < 10$ dB, on average, the proposed estimator and detector converge after $t_\text{ECM}=3$ and $t_\text{DATA DET}=5$ iterations, respectively, (ii) the number of iterations decreases to $t_\text{ECM}=t_\text{DATA DET}=2$ at $SNR \geq 20$ dB for $\sigma^2_\delta= 10^{-4} \ \text{rad}^2$ and $SNR \geq 30$ dB for $\sigma^2_\delta= 10^{-3} \ \text{rad}^2$, and (iii) the proposed ECM algorithm converges to the true estimates when the CFO estimates are initialized with a step size of $10^{-2}$, i.e., $t_\text{initialize} = 10^2$. \emph{For the existing algorithms}, the results in Section VI indicate that in order to reach an appropriate estimation accuracy and system performance, the algorithm in \cite{Lin-2006sept} requires the step size for the exhaustive search to be set to $10^{-3}$, i.e., $t_{[8]}$=$10^3$. In addition, the data detector in \cite{Lin-2007May} requires $t_{[20]} = 4$ iterations to converge for the PHN variance of $\sigma^2_\delta= [10^{-3}, 10^{-4}] \ \text{rad}^2$. Using these values for the number of iterations, we get the results shown in Fig. 3. We can see that the proposed estimation and data detection algorithms are computationally more efficient compared to the combined algorithms in \cite{Lin-2006sept} and \cite{Lin-2007May}, e.g., by a factor of $23.8$ for \mbox{SNR $= 20$ dB}, $\sigma^2_\delta= 10^{-4} \ \text{rad}^2$, $L=4$ and $N=64$.

\section{Simulation Results and Discussions} \label{sec-sec5}
In this subsection, we present simulation results to evaluate the performance of the proposed estimation and data detection algorithms.
We consider an OFDM packet to consist of $m=6$ OFDM symbols, comprising an OFDM training symbol followed by $5$ data symbols. The data symbols are drawn from normalized 64, 128, or 256 quadrature amplitude modulation (QAM). In the simulations, the symbol SNR is defined as $\xi_d/\sigma^2_w = 1/\sigma^2_w$. The sampling rate of the OFDM signal is $20$ MHz, corresponding to the OFDM sampling duration of $T_s = 50$ nanoseconds. The channel impluse response (CIR) is assumed to be a Rayleigh fading multipath channel with a delay of $L=4$ taps and an exponentially decreasing power delay profile with the average channel power \mbox{$= [-1.52 -6.75 -11.91 -17.08]$} dB. The Wiener PHN is generated with PHN variances of $\sigma^2_\delta=[10^{-3}, 10^{-4}]$ $\text{rad}^2$. The unknown normalized CFO is assumed to be uniformly distributed over the range $\epsilon \in (-0.5,0.5)$ for each simulation. Unless specified otherwise, an OFDM training symbol size of $N=64$ subcarriers is used with each subcarrier modulated in \emph{quadrature phase-shift keying (QPSK)} with subcarrier spacing $= 312.5$ kHz. The simulation results are averaged over $1 \times 10^5$ Monte Carlo simulation runs.

\subsection{Estimation Performance}
In this subsection, we compare the performance of the proposed ECM estimator with the HCRB in Theorem 1 and the MAP estimator in \cite{Lin-2006sept}. Figs. \ref{fig-4}, \ref{fig-5}, and \ref{fig-6} plot the HCRB and the mean square error (MSE) for estimating the channel impluse response, carrier frequency offset, and PHN, respectively. The HCRB in (\ref{eq-eqb5HCRB}) is numerically evaluated for two different PHN variance, e.g., $\sigma^2_\delta=[10^{-3},10^{-4}]$ $\text{rad}^2$.
The following observations can be made from the figures:
\begin{enumerate}
\item The HCRB for PHN and CFO estimation and the estimator's MSE are dependent on the variance of the PHN process. Note that, $\sigma^2_\delta=10^{-3}$ $\text{rad}^2$, corresponds to the presence of a very strong PHN \cite{Hani-2012sept}.
\begin{figure}[t]
  \begin{center}
    \includegraphics[width=0.53 \textwidth]{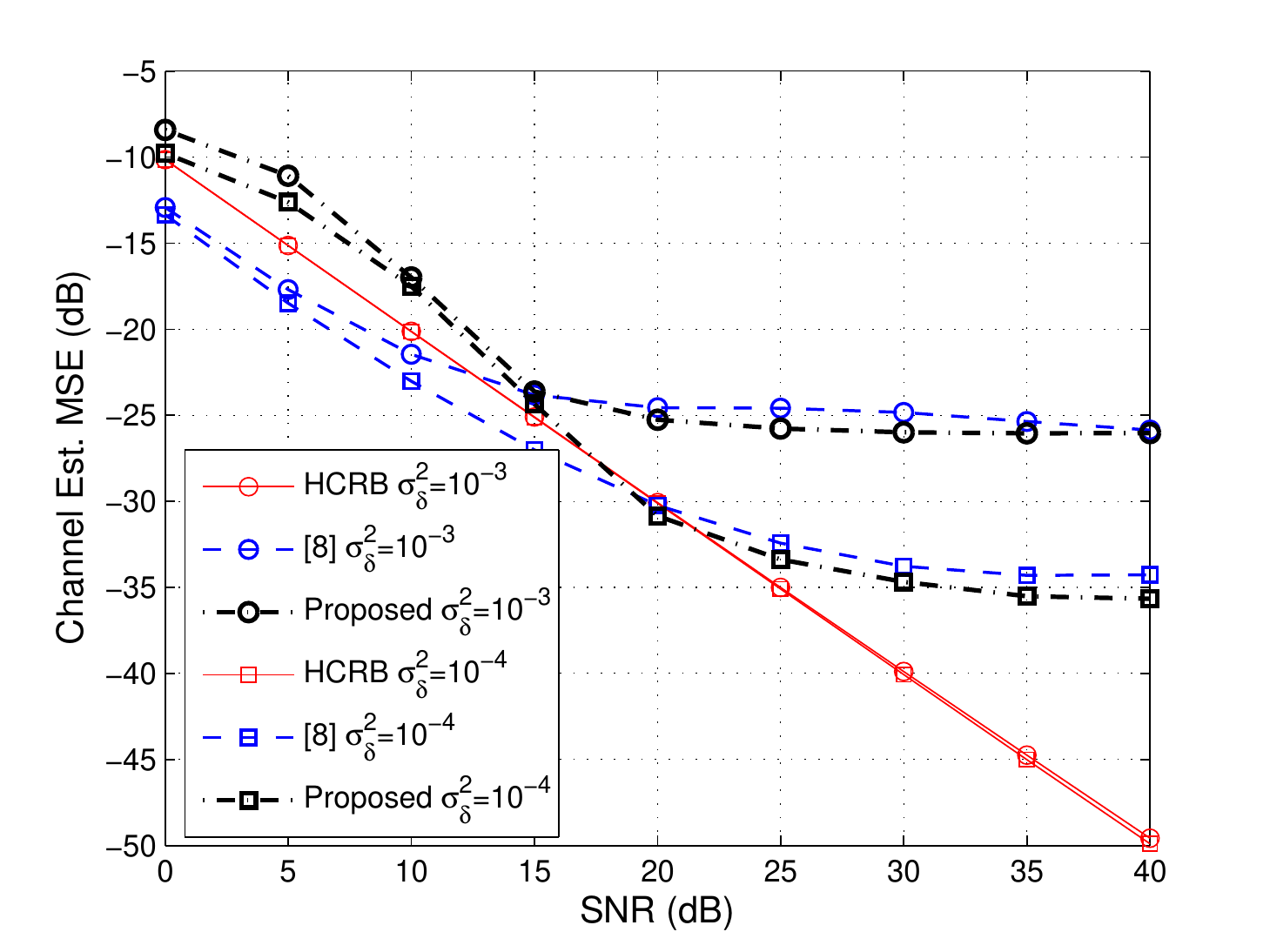}
   \end{center}
  \caption{Channel estimation MSE for the proposed and MAP estimators for PHN variance, $\sigma^2_\delta=[10^{-3}, 10^{-4}]$ $\text{rad}^2$.}
  \label{fig-4}
\end{figure}
\begin{figure*}[t]
     \begin{minipage}[h]{0.50\textwidth}
        \centering
    \includegraphics[width=1.05 \textwidth]{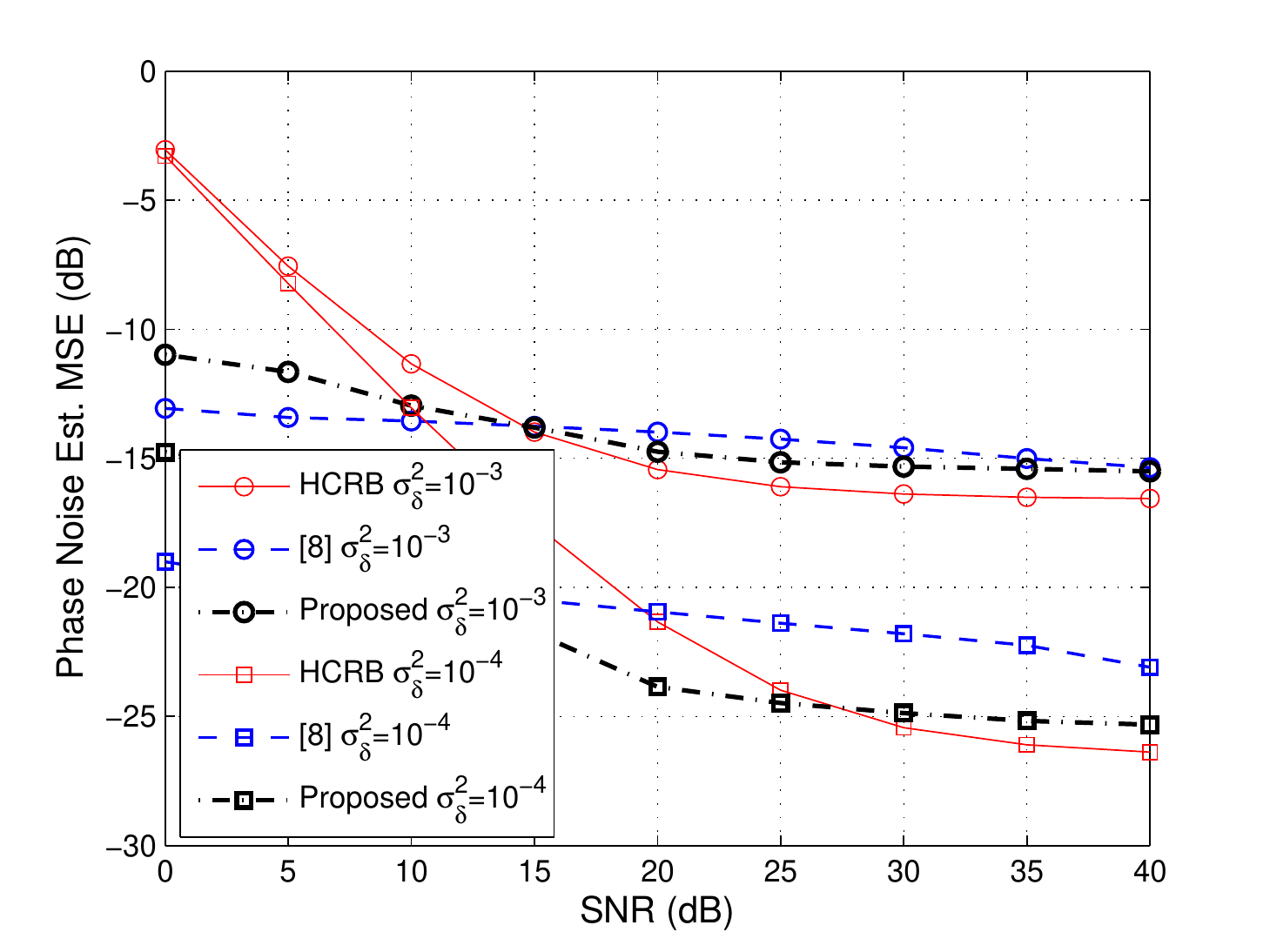}
    \caption{PHN estimation MSE for the proposed and MAP estimators for PHN variance, $\sigma^2_\delta=[10^{-3}, 10^{-4}]$ $\text{rad}^2$.}\label{fig-5}
    \end{minipage}
  \hspace{0.2in}
     \begin{minipage}[h]{0.50\textwidth}
        \centering
    \includegraphics[width=1.05 \textwidth]{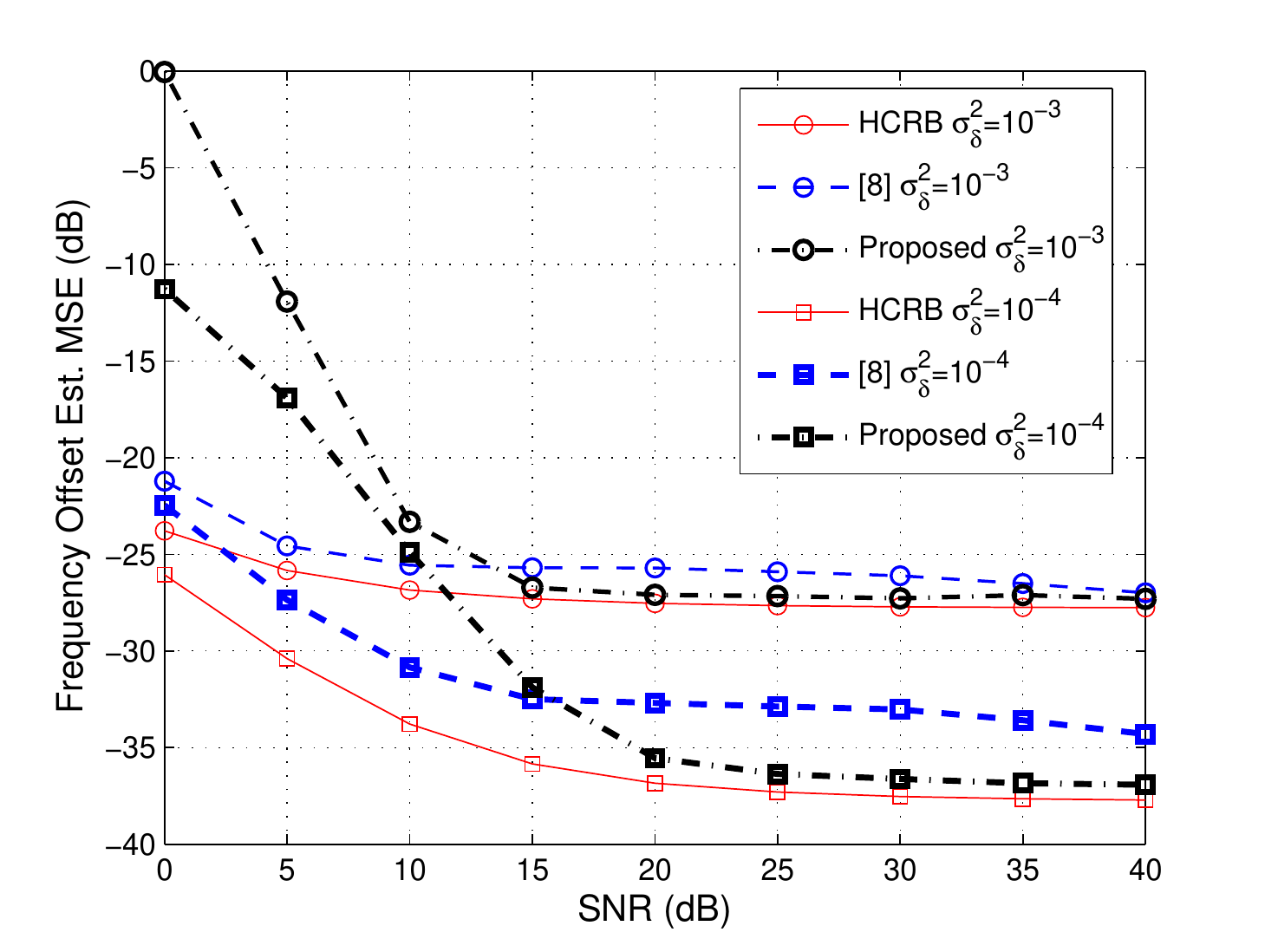}
  \caption{CFO estimation MSE for the proposed and MAP estimators for PHN variance, $\sigma^2_\delta=[10^{-3}, 10^{-4}]$ $\text{rad}^2$.}\label{fig-6}
  \end{minipage}
  \vspace{-0.18in}
\end{figure*}
\item Fig. \ref{fig-4} shows that the HCRB for the channel estimation does not suffer from an error floor, which is inline with Remark 1 in Section \ref{sec-sec2}. However, Figs. \ref{fig-5} and \ref{fig-6} show that the HCRB for CFO and PHN suffer from an error floor, which is directly related to the variance of the PHN process. This is due to the fact that at low SNR the performance of the system is dominated by AWGN, while at high SNR the performance of the proposed estimator is limited by PHN and the resulting ICI.
\item Figs. \ref{fig-4}, \ref{fig-5}, and \ref{fig-6} show that at low SNR, i.e., $SNR < 15$ dB, the proposed estimator is outperformed by the estimation algorithm in \cite{Lin-2006sept}. This outcome can be attributed to the different linearization approaches that are applied in both papers. In \cite{Lin-2006sept}, a first order Taylor series approximation is applied to linearize the signal model with respect to the PHN parameters over the whole OFDM symbol. However, the proposed algorithm uses an EKF algorithm that linearizes the observation sequence sample by sample, i.e., the estimate of the previous sample's PHN is used to linearize the current sample's PHN within the OFDM symbol. Thus, the proposed algorithm is less severely impacted by the residual error introduced by the first order Taylor series approximation. This results in significantly better estimation performance at high SNRs, where the estimator's performance is mainly impacted by PHN and not the AWGN. However, the estimator in \cite{Lin-2006sept} is based on the maximum a posterior (MAP) criterion that utilizes an exhaustive search to obtain the PHN parameters after applying the first order Taylor series approximation. Although this estimator is very complex, it is well-known that a MAP estimator is an optimal estimator for tracking random parameters and outperforms an EKF \cite{Nasir-2013Jan}. Hence, at low SNR, where the performance of the estimator is dictated by the AWGN noise, the MAP estimator in \cite{Lin-2006sept} is capable of more accurately tracking the PHN parameters compared to the EKF based estimator in this paper.

    It is important to note that, compared to \cite{Lin-2006sept}, the comparatively poor performance of the proposed estimator at low SNR does not result in significant degradation in the overall BER performance of an OFDM system (see Figs. \ref{fig-7}-\ref{fig-10}). This can be attributed to the fact that at low SNR the overall BER of the system is not dictated by the PHN estimation error but is bounded by the AWGN. However, at high SNR, where the overall performance of an OFDM system is PHN limited, the proposed algorithm demonstrates a significantly better BER performance due to the lower error associated with the estimation of PHN parameters compared to \cite{Lin-2006sept}.
\item Figs. \ref{fig-4}, \ref{fig-5}, and \ref{fig-6} show that the proposed estimator outperforms the estimation algorithm in \cite{Lin-2006sept} at moderate-to-high SNR. The biggest gain is achieved in the MSE of PHN estimation, followed by the MSE of CFO and channel estimation. For example, for $\sigma^2_\delta=10^{-4}$ $\text{rad}^2$ and at high SNR, the proposed estimator results in a $2$--$3$ dB performance gain compared to that of \cite{Lin-2006sept} while estimating PHN or CFO. This performance gain is in addition to the lower complexity of the proposed estimator as shown in Section \ref{sec-CA}.
\end{enumerate}

Note that in Fig. \ref{fig-5}, the PHN estimation MSE of the proposed estimator and the estimator in \cite{Lin-2007May} are lower than the HCRB at low SNR. This is due to the fact that the HCRB cannot be derived in closed-form while taking into account the prior knowledge of the range of CFO values, i.e., $(-0.5,0.5)$. However, the proposed estimator and the estimator in \cite{Lin-2007May} take into account this prior information while estimating the PHN, CFO, and the channel paraments.
\subsection{Comparison with Existing Work}
In the following, we examine the combined estimation and data detection performance in terms of the uncoded BER of the OFDM system. The following system setups are considered for comparison:

\noindent (i) The proposed estimation and data detection algorithm (labelled as ``Proposed Est. and Data Det.").

\noindent (ii) The estimation and data detection algorithm in \cite{Lin-2006sept} and \cite{Lin-2007May}, respectively (labelled as ``\cite{Lin-2006sept} $\&$ \cite{Lin-2007May}").

\noindent (iii) The data detection in \cite{Rabiei-20010Nov} combined with the proposed estimation algorithm (labelled as ``[Proposed estimation, 22] ").

\noindent (iv) The estimation algorithm in \cite{Lin-2006sept} combined with the proposed data detection algorithm (labelled as ``\cite[Proposed data detection]{Lin-2006sept} ").

\noindent (v) As a reference, a system that applies the proposed estimation algorithm but utilizes no PHN tracking during OFDM data symbols (labelled as ``No CFO cancel. and PHN track.").

\noindent (vi) As a lower-bound on the BER performance, a system assuming perfect channel, PHN, and CFO estimation (labelled as ``Perf. CIR, PHN \& CFO est.").

\begin{figure}[htp]
  \begin{center}
    \includegraphics[scale=0.53]{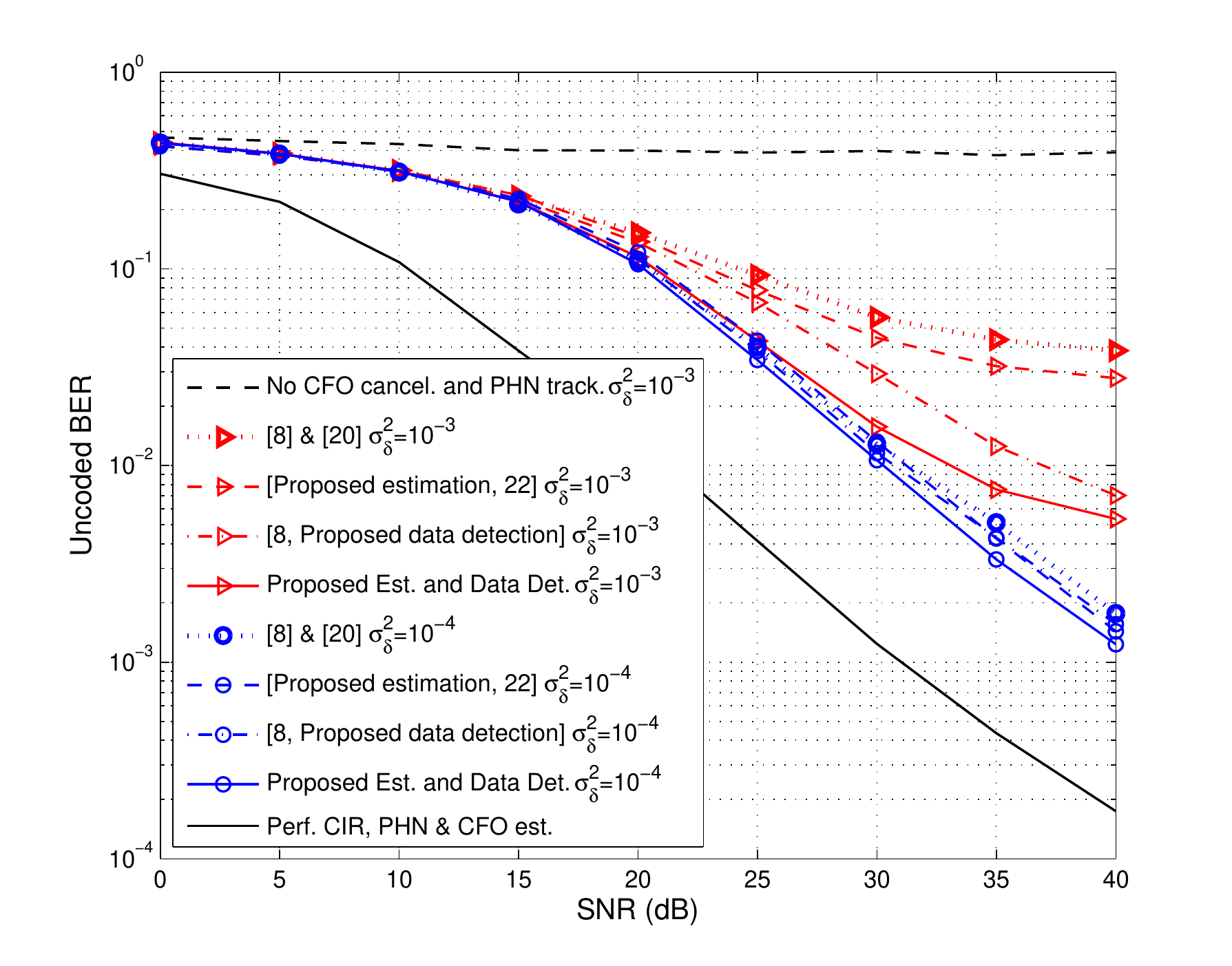}
   \end{center}
  \caption{Comparison of uncoded BER of the proposed algorithms for PHN variance, $\sigma^2_\delta= [10^{-3}, 10^{-4}] \ \text{rad}^2$ and 64-QAM modulation with the algorithms in \cite{Lin-2006sept} \& \cite{Lin-2007May} and \cite{Rabiei-20010Nov}.}
  \label{fig-7}
\end{figure}

Fig. \ref{fig-7} depicts the uncoded BER performances of the OFDM systems listed above. The following observations can be made from Fig. \ref{fig-7}:
\begin{enumerate}
\item The results demonstrate that without phase tracking and CFO cancellation throughout the packet, the OFDM system performance deteriorates significantly. On the other hand, by combining the proposed estimation and data detection algorithms, the BER performance of an OFDM system is shown to improve considerably even in the presence of very strong PHN, e.g., $\sigma^2_\delta=10^{-3}$ $\text{rad}^2$.
\item Compared to existing algorithms, the BER performance of an OFDM system using the proposed algorithms is closer to the ideal case of perfect CIR, PHN, and CFO estimation (a performance gap of 10 dB at SNR $=20$ dB).
\item It can be clearly observed that the proposed receiver structure outperforms the algorithms in \cite{Lin-2006sept} and \cite{Lin-2007May}. This performance improvement can be attributed to the fact that instead of single order Taylor series approximation applied directly to the whole OFDM symbol \cite{Lin-2006sept}, the proposed estimation and detection algorithms apply EKF to linearize the PHN, which uses the most recent estimated PHN values to obtain an updated PHN estimate sample by sample. This linearization using EKF helps in achieving better system performance at high SNR.
\item It is clear that the performance of the proposed data detection algorithm outperforms the algorithm in \cite{Rabiei-20010Nov}. This result is anticipated, since at high PHN variance, the approximation of PHN parameters using linear interpolation in \cite{Rabiei-20010Nov} highly deviates from the true PHN parameters. Therefore, the linear interpolation approach in \cite{Rabiei-20010Nov} may not be used in the presence of very strong PHN, e.g., $\sigma^2_\delta=10^{-3}$ $\text{rad}^2$.
\item It can be seen that the performance of the proposed estimator outperforms that of \cite{Lin-2006sept} even when the latter is combined with the proposed data detection algorithm. This is because the estimation based on EKF outperforms the estimation in \cite{Lin-2006sept} at high SNR. Thus, at high SNR, where the system performance is determined by the PHN estimation performance, our proposed estimation and detection algorithms achieve better BER results.
\item Finally, Fig. \ref{fig-7} shows that in the presence of PHN, the overall BER performance of an OFDM system suffers from an error floor at high SNR, since at high SNR the performance of an OFDM system is dominated by PHN, which cannot be completely eliminated.
\end{enumerate}

\begin{figure*}[t]
     \begin{minipage}[h]{0.50\textwidth}
        \centering
    \includegraphics[width=1.05 \textwidth]{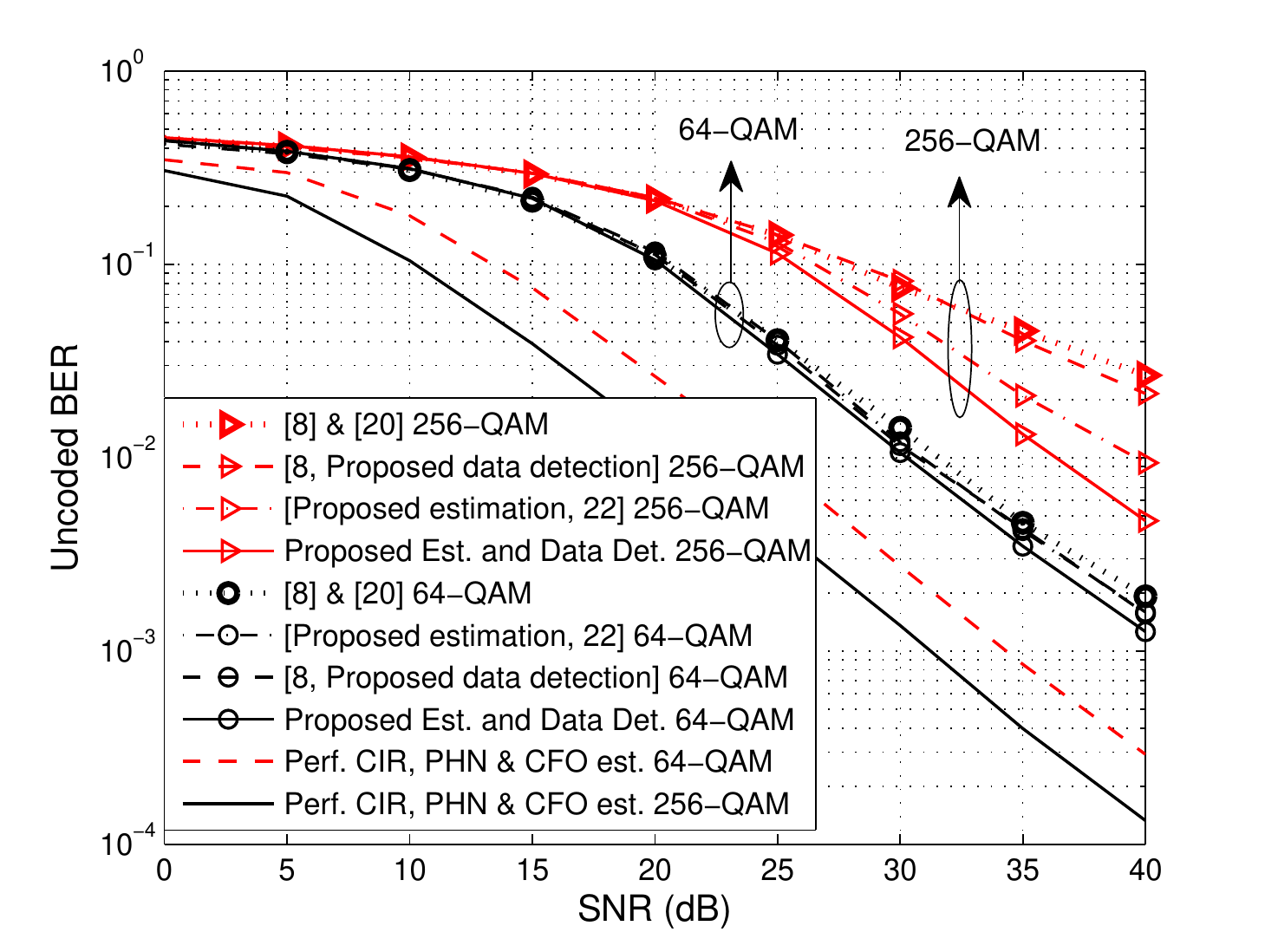}
    \caption{Comparison of uncoded BER of the proposed algorithms for 64-QAM and 256-QAM modulations with the algorithm in \cite{Lin-2006sept} \& \cite{Lin-2007May} and \cite{Rabiei-20010Nov} at PHN variance, $\sigma^2_\delta= 10^{-4} \ \text{rad}^2$.} \label{fig-9}
    \end{minipage}
  \hspace{0.2in}
     \begin{minipage}[h]{0.50\textwidth}
        \centering
    \includegraphics[width=1.05 \textwidth]{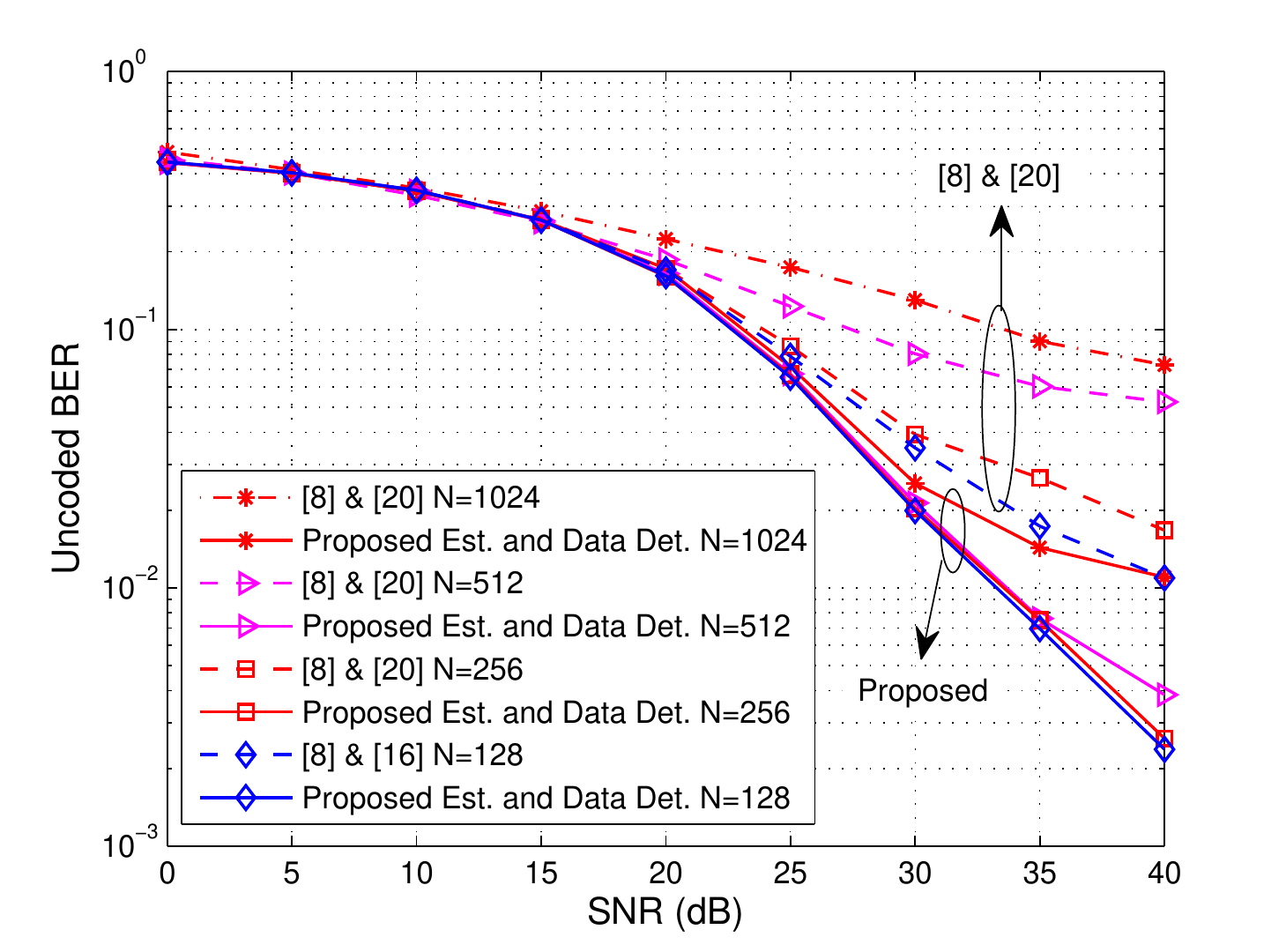}
  \caption{Comparison of uncoded BER of the proposed algorithms with the algorithm in \cite{Lin-2006sept} \& \cite{Lin-2007May} for varying training symbol lengths $N$= 128, 256, 512 and 1024, at $\sigma^2_\delta= 10^{-4} \ \text{rad}^2$ and 128-QAM modulation.} \label{fig-11}
  \end{minipage}
  \vspace{-0.18in}
\end{figure*}

\subsection{Effect of Modulation and OFDM System Parameter}

Fig. \ref{fig-9} illustrates the uncoded BER performance of an OFDM system for higher order modulations, i.e., 256-\emph{quadrature amplitude modulation} (256-QAM). The results in Fig. \ref{fig-9} shows that even for a  denser constellation, the proposed estimation and data detection algorithms significantly improve the overall system performance compared to that of [8, Proposed data detection], \cite{Lin-2007May} and \cite{Rabiei-20010Nov}. For example, to achieve a BER of $3 \times 10^{-2}$ with a PHN variance of $10^{-4} \ \text{rad}^2$, the proposed algorithm outperforms the algorithms in [8, Proposed data detection] and \cite{Lin-2007May} by a margin of $6$ dB and $7$ dB, respectively. In addition, the proposed algorithm outperforms the algorithm in \cite{Rabiei-20010Nov} by a margin of $3$ dB at a BER of $10^{-2}$ with a PHN variance of $10^{-4} \ \text{rad}^2$. In addition, this gap widens at higher SNR values.

Fig. \ref{fig-11} illustrates the uncoded BER performance of an OFDM system for different number of subcarriers, e.g., $N$ $= 128, 256, 512$ and $1024$, within an OFDM symbol. Based on the results in Fig. \ref{fig-11}, it can be concluded that the proposed algorithm is not sensitive to the subcarrier spacing at low-to-medium SNRs while at high SNRs, the BER degrades as one moves to $N=1024$ subcarriers. This is because increasing the number of subcarriers results in more ICI that is caused by the residual PHN and CFO. More importantly, the BER performance of an OFDM system using the proposed algorithms outperforms that of \cite{Lin-2006sept} and \cite{Lin-2007May} for any value of $N$. For instance, at BER = $10^{-2}$ and $N = 128$, the SNR gain for the proposed algorithms is almost 8 dB compared to the algorithms in \cite{Lin-2006sept} and \cite{Lin-2007May}.

\begin{figure}[htp]
  \begin{center}
    \includegraphics[scale=0.7]{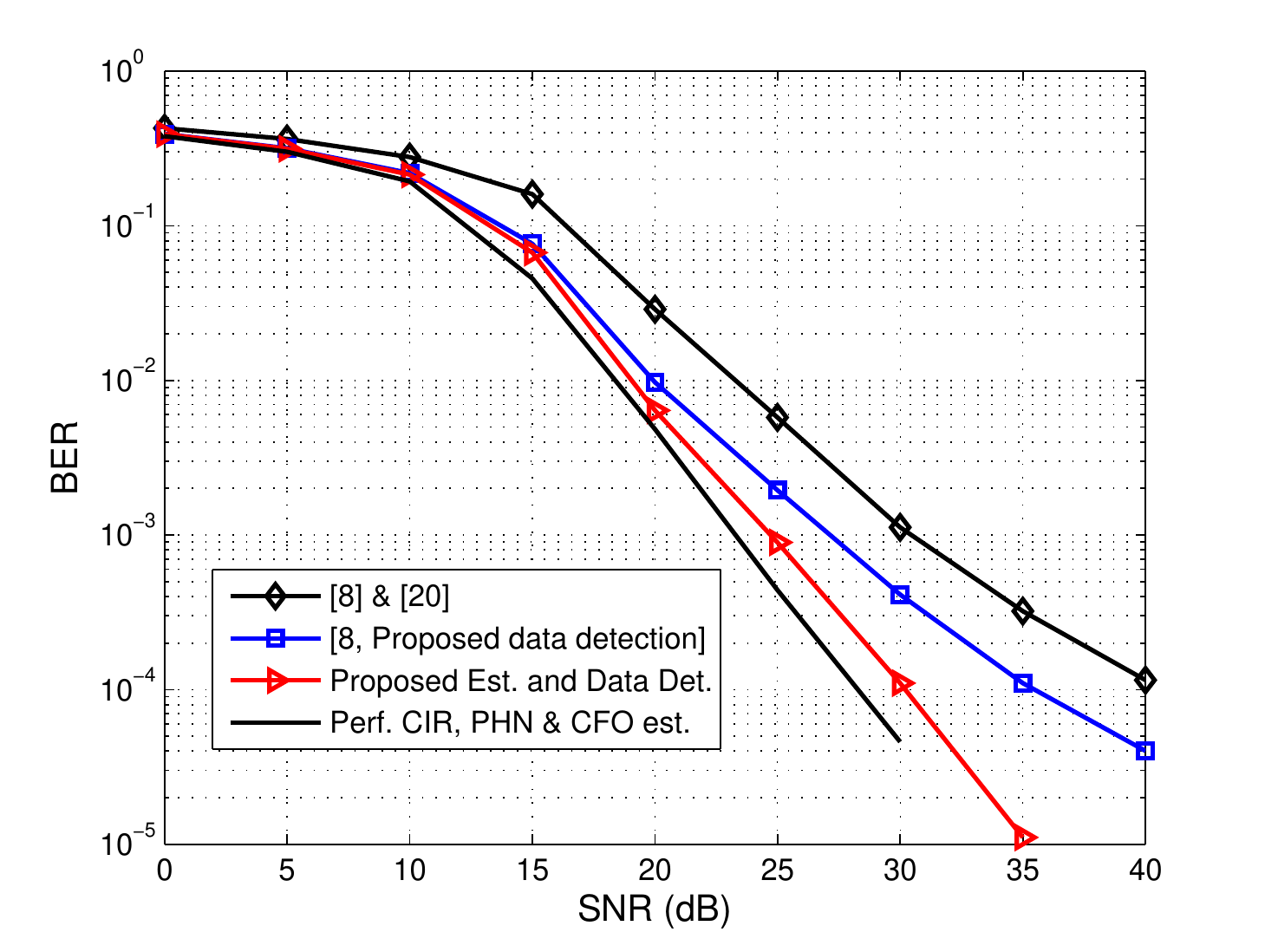}
   \end{center}
  \caption{Comparison of coded BER of the proposed algorithms with the algorithms in \cite{Lin-2006sept} \& \cite{Lin-2007May} and [8, Proposed data detection] for PHN variance, $\sigma^2_\delta= 10^{-4} \ \text{rad}^2$ and 64-QAM modulation.}
  \label{fig-10}
\end{figure}

Finally, the coded BER performance is shown in Fig. \ref{fig-10}. A low-density parity-check (LDPC) code is employed with a channel coding rate of $1/2$ and codeword length of $1296$ bits. The algorithm in \cite{Thomas-2001Feb} is used for encoding. The soft-decision iterative decoding algorithm, based on a sum-product algorithm in \cite{Johnson-Introducing} is utilized for decoding the estimated data vector in \eqref{eq-eq54}. The results in Fig. \ref{fig-10} show that, the proposed estimation and data detection algorithms improve the overall system performance compared to the existing algorithms in \cite{Lin-2006sept} and \cite{Lin-2007May}. For example, to achieve a BER of $10^{-4}$ with a PHN variance of $10^{-4}$ rad$^2$, the proposed algorithm outperforms the algorithms in ``[8, Proposed data detection]" and ``\cite{Lin-2006sept} \& \cite{Lin-2007May}" by a margin of $5$ dB and $10$ dB, respectively. In addition, compared to existing algorithms, the coded BER performance of an OFDM system using the proposed algorithm is closer to that of the ideal case of perfect CIR, PHN, and CFO estimation (a performance gap of $2$ dB at BER of $10^{-4}$).

\section{Conclusion} \label{sec-sec6}
In this paper, an ECM based algorithm has been proposed for joint estimation of channel, PHN, and CFO in OFDM systems. The signal model for the estimation problem is outlined in detail and the HCRB for the joint estimation of channel, PHN, and CFO in OFDM systems is derived. Simulation results indicate that the estimation MSE of the proposed algorithm is closer to the derived HCRB and outperforms the existing estimation algorithms at moderate-to-high SNR. Next, an \emph{iterative} algorithm for joint data detection and PHN mitigation is proposed for the OFDM data symbols. The proposed algorithm employs an EKF based approach to track the time-varying PHN parameters throughout the OFDM data symbols. The performance of the proposed estimation and detection algorithms has been evaluated for different PHN variances $\sigma^2_\delta= [10^{-3} , 10^{-4}]\text{rad}^2$, different number of subcarriers $N = [64, 128, 256, 512,1024]$, and different modulation schemes, $64,128,256$-QAM. Numerical results show that the proposed ECM based estimator and the iterative data detection algorithm are not only computationally efficient compared to the existing algorithms but also outperform existing algorithms in terms of both the uncoded and the coded BER performance. For example, the uncoded BER for the proposed algorithms has an SNR gain of almost $8$ dB compared to the existing algorithms at an BER of $10^{-2}$ with $N=128$. In addition, the coded BER performance using the proposed algorithms is closer to the ideal case of perfect CIR, PHN, and CFO estimation with a performance gap of only $2$ dB at BER of $10^{-4}$ and PHN variance, $\sigma^2_\delta= 10^{-4} \ \text{rad}^2$.

It is worth mentioning that the proposed estimation and detection algorithms in this paper can be modified for application in multi-input multi-output (MIMO) systems, since similar to single-input single-output systems, MIMO systems are also affected by a multiplicative phase noise factor and a CFO. However, addressing this specific problem is outside the scope of this paper and can be the subject of future work. Another open research problem is to derive the theoretical BER expression for an OFDM system in the presence of CFO, PHN and channel estimation.

\appendices
\numberwithin{equation}{section}

\section{Derivation of the HCRB} \label{sec-Jd}
The hybrid information matrix $\mathbf{B}$ can be written as \cite[pp. 1-85]{Trees-2007USA}
\begin{equation} \label{eq-eqb6}
\mathbf{B}=\boldsymbol{\Xi}_D+\boldsymbol{\Xi}_P,
\end{equation}
where $\boldsymbol{\Xi}_D \triangleq \mathbb{E}_{\boldsymbol{\theta}}\left[\boldsymbol{\Psi}(\boldsymbol{\theta},\Re\{\mathbf{h}\},\Im\{\mathbf{h}\},\epsilon)\right]$ with $\boldsymbol{\Psi}(\boldsymbol{\theta},\Re\{\mathbf{h}\},\Im\{\mathbf{h}\},\epsilon) \triangleq \mathbb{E}_{\mathbf{r}|\boldsymbol{\theta},\Re\{\mathbf{h}\}, \Im\{\mathbf{h}\},\epsilon}\big[- \Delta_{\boldsymbol{\lambda}}^{\boldsymbol{\lambda}} \log p(\mathbf{r}|\boldsymbol{\theta},\Re\{\mathbf{h}\},$ $\Im\{\mathbf{h}\},\epsilon)|,\Re\{\mathbf{h}\},\Im\{\mathbf{h}\},\epsilon\big]$ denoting the Fisher's information matrix (FIM) and $\boldsymbol{\Xi}_P \triangleq \mathbb{E}_{\boldsymbol{\theta}|,\Re\{\mathbf{h}\},\Im\{\mathbf{h}\},\epsilon}\big[- \Delta_{\boldsymbol{\lambda}}^{\boldsymbol{\lambda}} \log$ $p(\boldsymbol{\theta}|,\Re\{\mathbf{h}\},\Im\{\mathbf{h}\},\epsilon)|,\epsilon\big]$ is the prior information matrix with $p(\boldsymbol{\theta|\mathbf{h},\epsilon})$ denoting the prior distribution of PHN vector given the CIR and CFO. Thus, we first obtain expressions for matrices $\boldsymbol{\Xi}_D$ and $\boldsymbol{\Xi}_P$.

\subsection{Computation of $\boldsymbol{\Xi}_D \triangleq \mathbb{E}_{\boldsymbol{\theta}}\left[\boldsymbol{\Psi}(\boldsymbol{\theta},\Re\{\mathbf{h}\},\Im\{\mathbf{h}\},\epsilon)\right]$}

To compute FIM, first, the likelihood function $p(\mathbf{r}|\boldsymbol{\theta},\Re\{\mathbf{h}\},$ $\Im\{\mathbf{h}\},\epsilon)$ is given by
\begin{equation} \label{eq-eqb7}
p(\mathbf{r}|\boldsymbol{\theta},\Re\{\mathbf{h}\},\Im\{\mathbf{h}\},\epsilon)= \text{C} \ \text{exp} \left[\frac{-1}{\sigma^2_w}(\mathbf{r}-\boldsymbol{\mu})^H(\mathbf{r}-\boldsymbol{\mu})\right],
\end{equation}
where $\text{C} \triangleq (\pi \sigma^2_w)^{-N}$. Given $\boldsymbol{\theta}$, $\Re\{\mathbf{h}\}$, $\Im\{\mathbf{h}\}$, and $\epsilon$, $\mathbf{r}$ is a complex Gaussian vector with mean vector $\boldsymbol{\mu}=\mathbf{E}\mathbf{P}\mathbf{F}^H\mathbf{D}\mathbf{W}\mathbf{h}$ and covariance matrix $\sigma^2_w\mathbf{I}_N$.
The FIM, $\boldsymbol{\Psi}(\boldsymbol{\theta},\Re\{\mathbf{h}\},\Im\{\mathbf{h}\},\epsilon)$, will be $(N+2L) \times (N+2L)$ matrix for joint estimation of $(N-1)$ PHN parameters $\boldsymbol{\theta}$, $2L$ channel parameters $\Re\{\mathbf{h}\}$ and $\Im\{\mathbf{h}\}$ and one CFO parameter $\epsilon$. Using \eqref{eq-eqb7}, the $(i,j)$th entry of $\boldsymbol{\Psi}$ can be written as \cite{Kay-1993}
\begin{equation} \label{eq-eqb8}
[\boldsymbol{\Psi}]_{i,j}= \frac{2}{\sigma^2_w} \Re \left\{\frac{\partial \boldsymbol{\mu}^H}{\partial \boldsymbol{\lambda}_i} \frac{\partial \boldsymbol{\mu}}{\partial \boldsymbol{\lambda}_j} \right\},
\end{equation}
where
\begin{equation}
\label{eq-eqb9}
\frac{\partial \boldsymbol{\mu}^H}{\partial \boldsymbol{\lambda}_i}=\begin{cases}
\text{diag}(\mathbf{E}\mathbf{F}^H\mathbf{D}\mathbf{W}^H\mathbf{h})\mathbf{a}_i, & i=1,\ldots,N-1 \ \ (\boldsymbol{\lambda}_i=\theta_i)\\
\mathbf{E}\mathbf{P}\mathbf{F}^H\mathbf{D}\mathbf{W}^H  \mathbf{e}_{i-(N-1)}, & i=N,\ldots,N+L-1 \ \ (\boldsymbol{\lambda}_i=\Re\{h_{i-N}\})\\
j \mathbf{E}\mathbf{P}\mathbf{F}^H\mathbf{D}\mathbf{W}^H \mathbf{e}_{i-(N+L-1)}, & i=N+L,\ldots,N+2L-1 \ \ (\boldsymbol{\lambda}_i=\Im\{h_{i-(N+L)}\})\\
j \sqrt{\mathbf{M}}\mathbf{E}\mathbf{P}\mathbf{F}^H\mathbf{D}\mathbf{W}^H\mathbf{h}, & i=N+2L \ \ (\boldsymbol{\lambda}_i=\epsilon)
\end{cases}
\end{equation}
$\mathbf{a}_i=[0,0_{1 \times i-1},j e^{j \theta_i},0_{1 \times N-i}]^T$ for $i = 1,\ldots,N-1$, $\mathbf{e}_l=[0_{1 \times l-1},1,0_{1 \times L-1}]^T$ for $l=1,\ldots,L$, and $\mathbf{M} \triangleq \text{diag} \big(\big[(2 \pi \frac{0}{N})^2, (2 \pi \frac{1}{N})^2,$ $ \ldots,(2 \pi \frac{N-1}{N})^2\big]^T\big)$.

Substituting \eqref{eq-eqb9} into \eqref{eq-eqb8}, the matrix $\boldsymbol{\Xi}_D$ is obtained as
\begin{align} \label{eq-eqb8}
\boldsymbol{\Xi}_D&=\frac{2}{\sigma^2_w}\Re\left\{\left[
                                         \begin{array}{cccc}
                                         \bar{\mathbf{Q}}_1^H \bar{\mathbf{Q}}_1 & -j \bar{\mathbf{Q}}_1^H \bar{\mathbf{Q}}_2 & \bar{\mathbf{Q}}_1^H \bar{\mathbf{Q}}_2 & \bar{\mathbf{Q}}_4^H \bar{\mathbf{q}}_3 \\
                                         j \bar{\mathbf{Q}}_2^H \bar{\mathbf{Q}}_1 & \mathbf{Q}_2^H \mathbf{Q}_2 & j \mathbf{Q}_2^H \mathbf{Q}_2 & j \mathbf{Q}_2^H \mathbf{q}_5 \\
                                         \bar{\mathbf{Q}}_2^H \bar{\mathbf{Q}}_1 & -j \mathbf{Q}_2^H \mathbf{Q}_2 & \mathbf{Q}_2^H \mathbf{Q}_2 & \mathbf{Q}_2^H \mathbf{q}_5 \\
                                         \bar{\mathbf{q}}_3^H \bar{\mathbf{Q}}_4 & -j \bar{\mathbf{q}}_5^H \mathbf{Q}_2 & \mathbf{q}_5^H \mathbf{Q}_2 & \mathbf{q}_5^H \mathbf{q}_5 \\
                                         \end{array}
                                         \right]\right\}, \nonumber\\
\end{align}
where $\mathbf{Q}_1 = \text{diag}(\mathbf{F}^H\mathbf{D}\mathbf{W}\mathbf{h})$, $\bar{\mathbf{Q}}_1 = \mathbf{Q}_1(2:N,2:N)$, $\mathbf{Q}_2 = \mathbf{F}^H\mathbf{D}\mathbf{W}$, $\bar{\mathbf{Q}}_2 = \mathbf{Q}_2(2:N,1:L)$, $\mathbf{Q}_4 = \text{diag}(\sqrt{\mathbf{M}}\mathbf{F}^H\mathbf{D}\mathbf{W}\mathbf{h})$, $\bar{\mathbf{Q}}_4 = \mathbf{Q}_4(2:N,2:N)$, $\mathbf{q}_3 = \mathbf{F}^H\mathbf{D}\mathbf{W}\mathbf{h}$, and $\bar{\mathbf{q}}_3 = \mathbf{q}_3(2:N)$, and $\mathbf{q}_5 = \sqrt{\mathbf{M}}\mathbf{F}^H\mathbf{D}\mathbf{W}\mathbf{h}$. Note that the elements of $\boldsymbol{\theta}$ get canceled by their conjugates, hence, there is no need to calculate the explicit expectation of $\boldsymbol{\Psi}$ over $\boldsymbol{\theta}$.

\noindent
\subsection{Computation of $\boldsymbol{\Xi}_P \triangleq \mathbb{E}_{\boldsymbol{\theta}|\mathbf{h},\epsilon}\left[- \Delta_{\boldsymbol{\lambda}}^{\boldsymbol{\lambda}} \log p(\boldsymbol{\theta}|\mathbf{h},\epsilon)|\mathbf{h},\epsilon\right]$} \label{sec-Jp}
The second factor in HIM, defined in \eqref{eq-eqb6}, can be written as:
\begin{align} \label{eq-eqb38}
\boldsymbol{\Xi}_P=\lefteqn{\mathbb{E}_{\boldsymbol{\theta}|\mathbf{h},\epsilon}\left[- \Delta_{\boldsymbol{\lambda}}^{\boldsymbol{\lambda}} \log p(\boldsymbol{\theta}|\mathbf{h},\epsilon)|{\epsilon}\right] \triangleq \left[
                                         \begin{array}{cccc}
                                         \boldsymbol{\Xi}_{P_{11}} & \boldsymbol{\Xi}_{P_{12}} & \boldsymbol{\Xi}_{P_{13}} & \boldsymbol{\xi}_{P_{14}} \\
                                         \boldsymbol{\Xi}_{P_{21}} & \boldsymbol{\Xi}_{P_{22}} & \boldsymbol{\Xi}_{P_{23}} & \boldsymbol{\xi}_{P_{24}} \\
                                         \boldsymbol{\Xi}_{P_{31}} & \boldsymbol{\Xi}_{P_{32}} & \boldsymbol{\Xi}_{P_{33}} & \boldsymbol{\xi}_{P_{34}} \\
                                         \boldsymbol{\xi}_{P_{41}} & \boldsymbol{\xi}_{P_{42}} & \boldsymbol{\xi}_{P_{43}} & \boldsymbol{\xi}_{P_{44}} \\
                                         \end{array}
                                         \right] } \nonumber\\
                                         & = \left[
                                         \begin{array}{cccc}
                                         \mathbb{E}_{\boldsymbol{\theta}}\left[- \Delta_{\boldsymbol{\theta}}^{\boldsymbol{\theta}} \log p(\boldsymbol{\theta})\right] &
                                         \mathbb{E}_{\boldsymbol{\theta}}\left[- \Delta_{\boldsymbol{\theta}}^{\Re\{\mathbf{h}\}} \log p(\boldsymbol{\theta})\right] &
                                         \mathbb{E}_{\boldsymbol{\theta}}\left[- \Delta_{\boldsymbol{\theta}}^{\Im\{\mathbf{h}\}} \log p(\boldsymbol{\theta})\right] &
                                         \mathbb{E}_{\boldsymbol{\theta}}\left[- \Delta_{\boldsymbol{\theta}}^{\epsilon} \log p(\boldsymbol{\theta})\right]\\
                                         \mathbb{E}_{\boldsymbol{\theta}}\left[- \Delta_{\Re\{\mathbf{h}\}}^{\boldsymbol{\theta}} \log p(\boldsymbol{\theta})\right] &
                                         \mathbb{E}_{\boldsymbol{\theta}}\left[- \Delta_{\Re\{\mathbf{h}\}}^{\Re\{\mathbf{h}\}} \log p(\boldsymbol{\theta})\right] &
                                         \mathbb{E}_{\boldsymbol{\theta}}\left[- \Delta_{\Re\{\mathbf{h}\}}^{\Im\{\mathbf{h}\}} \log p(\boldsymbol{\theta})\right] &
                                         \mathbb{E}_{\boldsymbol{\theta}}\left[- \Delta_{\Re\{\mathbf{h}\}}^{\epsilon} \log p(\boldsymbol{\theta})\right]\\
                                         \mathbb{E}_{\boldsymbol{\theta}}\left[- \Delta_{\Im\{\mathbf{h}\}}^{\boldsymbol{\theta}} \log p(\boldsymbol{\theta})\right] &
                                         \mathbb{E}_{\boldsymbol{\theta}}\left[- \Delta_{\Im\{\mathbf{h}\}}^{\Re\{\mathbf{h}\}} \log p(\boldsymbol{\theta})\right] &
                                         \mathbb{E}_{\boldsymbol{\theta}}\left[- \Delta_{\Im\{\mathbf{h}\}}^{\Im\{\mathbf{h}\}} \log p(\boldsymbol{\theta})\right] &
                                         \mathbb{E}_{\boldsymbol{\theta}}\left[- \Delta_{\Im\{\mathbf{h}\}}^{\epsilon} \log p(\boldsymbol{\theta})\right] \\
                                         \mathbb{E}_{\boldsymbol{\theta}}\left[- \Delta_{\epsilon}^{\boldsymbol{\theta}} \log p(\boldsymbol{\theta})\right] &
                                         \mathbb{E}_{\boldsymbol{\theta}}\left[- \Delta_{\epsilon}^{\Re\{\mathbf{h}\}} \log p(\boldsymbol{\theta})\right] &
                                         \mathbb{E}_{\boldsymbol{\theta}}\left[- \Delta_{\epsilon}^{\Im\{\mathbf{h}\}} \log p(\boldsymbol{\theta})\right] &
                                         \mathbb{E}_{\boldsymbol{\theta}}\left[- \Delta_{\epsilon}^{\epsilon} \log p(\boldsymbol{\theta})\right] \\
                                         \end{array}
                                         \right],
\end{align}
where $\triangle_{\textbf{y}}^{\mathbf{x}}$ represents the second-order partial derivative operator and $p(\boldsymbol{\theta})$ is the prior distribution of $\boldsymbol{\theta}$.

\noindent \emph{1) Computation of $ \boldsymbol\Xi_{P_{11}} \triangleq \mathbb{E}_{\boldsymbol{\theta}}\left[- \Delta_{\boldsymbol{\theta}}^{\boldsymbol{\theta}} \log p(\boldsymbol{\theta})\right]$}:
From \cite[eq.(19)]{Bay-2008Jan}, we obtain the  $(N-1) \times (N-1)$ matrix $\mathbb{E}_{\boldsymbol{\theta}}\left[- \Delta_{\boldsymbol{\theta}}^{\boldsymbol{\theta}} \log p(\boldsymbol{\theta})\right]$ as
\begin{equation}\label{eq-eqb39}
\boldsymbol{\Xi}_{P_{11}} =  \frac{-1}{\sigma^2_\delta} \left[ {\begin{array}{ccccc}
 -1 & 1 & 0 & \cdots & 0  \vspace{-0mm} \\
 1 & -2 & 1 & 0 & \vdots \vspace{-0mm} \\
  0 & \ddots & \ddots & \ddots & 0 \vspace{-0mm} \\
  \vdots & 0 & 1  & -2 & 1 \vspace{-0mm} \\
 0 & \cdots & 0 & 1 & -1
\end{array}  } \right].
\end{equation}

\noindent \emph{2) Computation of remaining terms in \eqref{eq-eqb38}}: Since CFO is a deterministic parameter and no prior knowledge of $\mathbf{h}$ is assumed, we have
\begin{align} \label{eq-eqb40}
&\boldsymbol{\Xi}_{P_{12}}=\boldsymbol{\Xi}_{P_{21}}^T=\mathbf{0}_{(N-1) \times L}, \\
&\boldsymbol{\Xi}_{P_{13}}=\boldsymbol{\Xi}_{P_{31}}^T=\mathbf{0}_{(N-1) \times L}, \\
&\boldsymbol{\Xi}_{P_{22}}=\boldsymbol{\Xi}_{P_{33}}=\boldsymbol{\Xi}_{P_{23}}=\boldsymbol{\Xi}_{P_{32}}=\mathbf{0}_{L \times L}, \\
&\boldsymbol{\xi}_{P_{14}}=\boldsymbol{\xi}_{P_{14}}^T=\mathbf{0}_{(N-1) \times 1}, \\
&\boldsymbol{\xi}_{P_{23}}=\boldsymbol{\xi}_{P_{32}}^T=\mathbf{0}_{L \times 1}, \\
&\boldsymbol{\xi}_{P_{24}}=\boldsymbol{\xi}_{P_{42}}^T=\boldsymbol{\xi}_{P_{34}}=\boldsymbol{\xi}_{P_{43}}^T=\mathbf{0}_{L \times 1}, \\
&\xi_{P_{44}}=0.
\end{align}
Using the above results, we can evaluate the HIM in \eqref{eq-eqb5HCRB}, since $\mathbf{B}_{11} = \boldsymbol{\Xi}_{D_{11}}+\boldsymbol{\Xi}_{P_{11}}$,
 $\mathbf{B}_{22} =\boldsymbol{\Xi}_{D_{22}}+\boldsymbol{\Xi}_{P_{22}}=\boldsymbol{\Xi}_{D_{22}}$,
 $\mathbf{B}_{33} =\boldsymbol{\Xi}_{D_{33}}+\boldsymbol{\Xi}_{P_{33}}=\boldsymbol{\Xi}_{D_{33}}$,
 $\text{b}_{44}=\xi_{D_{44}}+\xi_{P_{44}}=\xi_{D_{33}}$,
 $\mathbf{B}_{12} =\mathbf{B}_{21}^H=\boldsymbol{\Xi}_{D_{12}}+\boldsymbol{\Xi}_{P_{12}}=\boldsymbol{\Xi}_{D_{12}}$,
 $\mathbf{B}_{13} =\mathbf{B}_{31}^H=\boldsymbol{\Xi}_{D_{13}}+\boldsymbol{\Xi}_{P_{13}}=\boldsymbol{\Xi}_{D_{13}}$,
 $\mathbf{B}_{23} =\mathbf{B}_{32}^H=\boldsymbol{\Xi}_{D_{23}}+\boldsymbol{\Xi}_{P_{23}}=\boldsymbol{\Xi}_{D_{23}}$,
  $\mathbf{b}_{14}=\mathbf{b}_{41}^H=\boldsymbol{\xi}_{D_{14}}+\boldsymbol{\xi}_{P_{14}}=\boldsymbol{\xi}_{D_{14}}$, $\mathbf{b}_{24}=\mathbf{b}_{24}^H=\boldsymbol{\xi}_{D_{24}}+\boldsymbol{\xi}_{P_{24}}=\boldsymbol{\xi}_{D_{24}}$, and
  $\mathbf{b}_{34}=\mathbf{b}_{43}^H=\boldsymbol{\xi}_{D_{34}}+\boldsymbol{\xi}_{P_{34}}=\boldsymbol{\xi}_{D_{34}}$.

\ifCLASSOPTIONpeerreview
\vspace{-0.5cm}
\else
\fi


\begin{thebibliography}{10}
\providecommand{\url}[1]{#1}
\csname url@samestyle\endcsname
\providecommand{\newblock}{\relax}
\providecommand{\bibinfo}[2]{#2}
\providecommand{\BIBentrySTDinterwordspacing}{\spaceskip=0pt\relax}
\providecommand{\BIBentryALTinterwordstretchfactor}{4}
\providecommand{\BIBentryALTinterwordspacing}{\spaceskip=\fontdimen2\font plus
\BIBentryALTinterwordstretchfactor\fontdimen3\font minus
  \fontdimen4\font\relax}
\providecommand{\BIBforeignlanguage}[2]{{%
\expandafter\ifx\csname l@#1\endcsname\relax
\typeout{** WARNING: IEEEtran.bst: No hyphenation pattern has been}%
\typeout{** loaded for the language `#1'. Using the pattern for}%
\typeout{** the default language instead.}%
\else
\language=\csname l@#1\endcsname
\fi
#2}}
\providecommand{\BIBdecl}{\relax}
\BIBdecl

\bibitem{Nuaymi-2007wiley}
\BIBentryALTinterwordspacing
L. Nuaymi, \emph{WiMAX: Technology for broadband wireless access}, New York, USA, John Wiley \& Sons, 2007.
\BIBentrySTDinterwordspacing

\bibitem{Zhang-2008wiley}
Y. Zhang and H. -H. Chen, \emph{Mobile WIMAX: Toward broadband wireless metropolitan area networks}, New York, USA, Taylor \& Francis Group, 2008.

\bibitem{Paul-2008FQ}
\BIBentryALTinterwordspacing
T. Paul and T. Ogunfunmi, ``Wireless {LAN} comes of age: Understanding the {IEEE} 802.11n amendment,'' \emph{IEEE Circuits Syst. Mag.}, vol.~8, no.~1,  pp. 28--54, 2008.
\BIBentrySTDinterwordspacing

\bibitem{Zhu-2011May}
X. Zhu, A. Doufexi, and T. Kocak, ``Throughput and coverage performance for {IEEE} 802.11ad millimeter-wave {WPANs},'' in \emph{Proc. IEEE VTC (Spring)}, Budapest, Hungary, May 2011, pp. 1--5.

\bibitem{Chevillat-1987Sep}
\BIBentryALTinterwordspacing
P. Chevillat and D. Maiwald and G. Ungerboeck, ``Rapid training of a voiceband data-modem receiver employing an equalizer with
fractional-{T} spaced coefficients,'' \emph{IEEE Commun. Mag.}, vol.~35, pp. 869--876, Sep. 1987.
\BIBentrySTDinterwordspacing

\bibitem{Schmidl-1997Dec}
T. Schmidl and D. Cox, ``Robust frequency and timing synchronization for {OFDM},'' \emph{IEEE Commun. Mag.}, vol.~45, no.~12,  pp. 1613--1621, Dec. 1997.

\bibitem{article-PHASE-N-MODEL-I}
A. Chorti and M. Brookes, ``A spectral model for {RF} oscillators with power-law phase noise,'' \emph{IEEE Trans. Circuits Syst.}, vol.~53, no.~9,  pp. 1989--1999, Sep. 2006.

\bibitem{Lin-2006sept}
\BIBentryALTinterwordspacing
D. D. Lin, R. Pacheco, T. J. Lim, and D. Hatzinakos, ``Joint estimation of channel response, frequency offset, and phase noise in
{OFDM},'' \emph{IEEE Trans. Signal Process.}, vol.~54, no.~9, pp. 3542--3554, Sept. 2006.
\BIBentrySTDinterwordspacing

\bibitem{Armada-1998Jan}
\BIBentryALTinterwordspacing
A. Armada and M. Calvo, ``Phase noise and sub-carrier spacing effects on the performance of an {OFDM} communication system,'' \emph{IEEE Commun. Lett.}, vol.~2, no.~1, pp. 11--13, Jun. 1998.
\BIBentrySTDinterwordspacing

\bibitem{Armada-2001Jun}
\BIBentryALTinterwordspacing
A. Armada and M. Calvo, ``Understanding the effects of phase noise in orthogonal frequency division multiplexing ({OFDM}),'' \emph{IEEE Trans. Broadcast.}, vol.~47, no.~2, pp. 153--159, Jun. 2001.
\BIBentrySTDinterwordspacing

\bibitem{Petrovi1-2003}
\BIBentryALTinterwordspacing
D. Petrovi, W. Rave, and G. Fettweis, ``Phase noise suppression in OFDM including intercarrier interference,'' in \emph{Proc. Int. OFDM Workshop}, 2003, pp. 219--224.
\BIBentrySTDinterwordspacing

\bibitem{Petrovi2-2003}
D. Petrovi, W. Rave, and G. Fettweis, ``Phase noise suppression in {OFDM} using {Kalman} filter,'' in \emph{Proc. WPMC}, 2003.

\bibitem{Petrovic-2007Aug}
D. Petrovi, W. Rave, and G. Fettweis, ``Effects of phase noise on {OFDM} systems with and without {PLL}: Characterization and
compensation,'' \emph{IEEE Trans. Commun.}, vol.~55, no.~8, pp. 1607–-1616, Aug. 2007.

\bibitem{article_mehr-comm-13}
H. Mehrpouyan, M. R. Khanzadi, M. Matthaiou, A. M. Sayeed, R. Schober, and Y. Hua, ``Improving bandwidth efficiency in E-band communication systems,'' \emph{IEEE Commun. Mag.}, , vol.~52, no.~3, pp. 121--128, Mar. 2014.

\bibitem{Lee-2006Jan}
J. -H. Lee, J. C. Han, and S. -C. Kim, ``Joint carrier frequency synchronization and channel estimation for {OFDM} systems via the {EM} algorithm,'' \emph{IEEE Trans. Veh. Technol.}, vol.~55, no.~1, pp. 167--172, Jan. 2006.

\bibitem{Tao-2009Oct}
J. Tao, J. Wu, and C. Xiao, ``Estimation of channel transfer function and carrier frequency offset for OFDM systems with phase noise,'' \emph{IEEE Trans. Veh. Technol.}, vol.~58, no.~8, pp. 4380–-4387, Oct. 2009.

\bibitem{Carvajal-2013Jan}
R. Carvajal, J. C. Aguero, B. I. Godoy, and G. C. Goodwin, ``EM-based maximum-likelihood channel estimation in multicarrier systems
with phase distortion,'' \emph{IEEE Trans. Veh. Technol.}, vol.~62, no.~1, pp. 152--160, Jan. 2013.

\bibitem{Septier-2008Aug}
F. Septier, Y. Delignon, A. Menhaj-Rivenq, and C. Garnier, ``Monte Carlo methods for channel, phase noise, and frequency offset estimation with unknown noise variances in OFDM systems,'' \emph{IEEE Trans. Signal Process.}, vol.~56, no.~8, pp. 3613–-3626, Aug. 2008.

\bibitem{article-rui-sp-2013}
R. Wang and H. Mehrpouyan and M. Tao and Y. Hua, ``Channel estimation, carrier recovery, and data detection in the presence
	of phase noise in OFDM relay systems,'' \emph{submitted to IEEE Trans. Signal Proc.} [Online]. Available: http://arxiv.org/pdf/1309.7564.pdf

\bibitem{Lin-2007May}
D. D. Lin and T. J. Lim, ``The variational inference approach to joint data detection and phase noise estimation in OFDM,'' \emph{IEEE Trans. Signal Process.}, vol.~55, no.~5, pp. 1862--1874, May 2007.

\bibitem{Gong-2008May}
Y. Gong and X. Hong, ``A new algorithm for OFDM joint data detection and phase noise cancellation,'' in \emph{Proc. IEEE ICC}, Beijing, China, May 2008, pp.636--640.

\bibitem{Rabiei-20010Nov}
P. Rabiei, W. Namgoong, and N. Al-Dhahir, ``A non-iterative technique for phase noise ICI mitigation in packet-based OFDM systems,'' \emph{IEEE Trans. Signal Process.}, vol.~58, no.~11, pp. 5945--5950, Nov. 2010.

\bibitem{Omar-2013June}
O. H. Salim, A. A. Nasir, H. Mehrpouyan, and W. Xiang, ``Phase noise and carrier frequency offset in {OFDM} systems: Joint estimation and hybrid {C}ram\'{e}r-{R}ao lower bound,'' in \emph{Proc. IEEE SPAWC}, Darmstadt, Germany, Jun. 2013.

\bibitem{Demir-2000May}
A. M. A. Demir and J. Roychowdhury, ``Phase noise in oscillators: A unifying theory and numerical methods for characterization,'' \emph{IEEE Trans. Biomed. Circuits Syst.}, vol.~47, pp. 655--674, May 2000.

\bibitem{Lin-2006apr}
D. Lin, R. Pacheco, T. J. Lim, and D. Hatzinakos, ``Optimal OFDM channel estimation with carrier frequency offset and phase noise,'' in \emph{Proc. IEEE WCNC}, Apr. 2006.

\bibitem{Wu-2002Dec}
S. Wu and Y. Bar-Ness, ``A phase noise suppression algorithm for OFDM-based WLANs,'' \emph{IEEE Commun. Lett.}, vol.~6, no.~12, pp. 535--537, Dec. 2002.

\bibitem{Munier-2008Jul}
F. Munier, T. Eriksson, and A. Svensson, ``An ICI reduction scheme for OFDM system with phase noise over fading channels,'' \emph{IEEE Trans. Commun.}, vol.~56, no.~12, pp. 1119--1126, Jul. 2008.

\bibitem{Trees-2007USA}
H. L. van Trees and K. L. Bell, \emph{Bayesian bounds for parameter estimation and nonlinear filtering/tracking.} New York: Wiley-Intersecience; Wiley; IEEE, 2007.

\bibitem{Simon-2012Feb}
E. P. Simon, L. Ros, H. Hijazi, and M. Ghogho, ``Joint carrier frequency offset and channel estimation for OFDM systems via the EM algorithm in the presence of very high mobility,'' \emph{IEEE Trans. Signal Process.}, vol.~60, no.~2, pp. 754--765, Feb. 2012.

\bibitem{Kay-1993}
S. M. Kay, \emph{Fundamentals of Statistical Signal Processing, Estimation Theory.} Signal Processing Series: Prentice Hall, 1993.

\bibitem{Meng-1993Meng}
X. L. Meng and D. B. Rubin, ``Maximum likelihood estimation via the ECM algorithm: A general framework,'' \emph{IEEE Trans. Commun.}, \emph{Biometrika}, vol.~8, no.~2, pp. 267--278, Jun. 1993.

\bibitem{McLachlan-2008Wiley}
G. J. McLachlan and T. Krishnan, \emph{The EM Algorithm and Extensions.} Hoboken, NJ: Wiley, 2008.

\bibitem{Fesler-94-A}
T. A. Fesler and A. O. Hero, ``Space-alternating generalized expectation maximization algorithm,'' \emph{IEEE Trans. Signal Process.}, vol.~42, no.~10, pp. 2664--2677, Oct. 1994.

\bibitem{Hani-2011Apr}
H. Mehrpouyan and S. D. Blostein, ``Bounds and algorithms for multiple frequency offset estimation in cooperative networks,'' \emph{IEEE Trans. Wireless Commun.}, vol.~10, no.~4, pp. 1300--1311, Apr. 2011.

\bibitem{Hani-2012sept}
H. Mehrpouyan, A. A. Nasir, S. D. Blostein, T. Eriksson, G. K. Karagiannidis, and T. Svensson, ``Joint estimation of channel and oscillator phase noise in {MIMO} systems,'' \emph{IEEE Trans. Signal Process.}, vol.~60, no.~9, pp. 4790–-4807, Sept. 2012.

\bibitem{Nasir-2013Jan}
A. A. Nasir and H. Mehrpouyan and R. Schober and Y. Hua, ``Phase Noise in {MIMO} Systems: {B}ayesian {C}ram\'{e}r-{R}ao Bounds and Soft-Input Estimation,'' \emph{IEEE Trans. Signal Process.}, vol.~61, no.~10, pp. 2675--2692, May, 2013.

\bibitem{Thomas-2001Feb}
T. J. Richardson and R. L. Urbanke, ``Efficient encoding of low-density parity-check codes,'' \emph{IEEE Trans. Inf. Theory}, vol.~47, no.~2, pp. 638 - 656, Feb. 2001.

\bibitem{Johnson-Introducing}
S. J. Johnson, ``Introducing low-density parity-check codes,'' [Online]. Available: http://sigpromu.org/sarah/SJohnsonLDPCintro.pdf

\bibitem{Bay-2008Jan}
S. Bay, C. Herzet, J. -M. Brossier, J. -P. Barbot, and B. Geller, ``Analytic and asymptotic analysis of Bayesian Cram´er-Rao bound for dynamical phase offset estimation,'' \emph{IEEE Trans. Signal Process.}, vol.~56, no.~1, pp. 61--70, Jan. 2008.


\end{thebibliography}
\end{document}